\documentclass[fleqn,usenatbib,useAMS]{mnras}

\usepackage{graphicx}	
\usepackage{amsmath}	
\usepackage{amssymb}	
\usepackage{multicol}        
\usepackage{bm}		
\usepackage{pdflscape}	

\usepackage{subcaption}

\usepackage{enumitem}

\usepackage{booktabs}
\usepackage{multirow}

\usepackage{stfloats}

\usepackage[T1]{fontenc}
\usepackage{ae,aecompl}

\usepackage{newtxtext,newtxmath}

	\title[State-space PTA]{State-space analysis of a continuous gravitational wave source with a pulsar timing array: inclusion of the pulsar terms}

\author[Kimpson]{Tom Kimpson$^{1,2}$\thanks{Contact e-mail: \href{tom.kimpson@unimelb.edu.au}{tom.kimpson@unimelb.edu.au}}, Andrew Melatos$^{1,2}$, Joseph O'Leary$^{1,2}$, Julian B. Carlin$^{1,2}$, Robin J. Evans$^{3}$, \newauthor William Moran$^{3}$, Tong Cheunchitra$^{1,2}$, Wenhao Dong$^{1,2}$, Liam Dunn$^{1,2}$, Julian Greentree$^{3}$, Nicholas J. O'Neill$^{1,2}$, \newauthor Sofia Suvorova$^{3}$, Kok Hong Thong$^{1,2}$, Andrés F. Vargas$^{1,2}$%
\\
$^{1}$School of Physics, University of Melbourne, Parkville, VIC 3010, Australia \\
$^{2}$OzGrav, University of Melbourne, Parkville, VIC 3010, Australia \\
$^{3}$Department of Electrical and Electronic Engineering, University of Melbourne, Parkville, VIC 3010, Australia }

\date{Last updated \today}

\pubyear{2024}	

\begin{document}
\label{firstpage}
\pagerange{\pageref{firstpage}--\pageref{lastpage}}
\maketitle

\begin{abstract}	
	 Pulsar timing arrays can detect continuous nanohertz gravitational waves emitted by individual supermassive black hole binaries. The data analysis procedure can be formulated within a time-domain, state-space framework, in which the radio timing observations are related to a temporal sequence of latent states, namely the intrinsic pulsar spin frequency. The achromatic wandering of the pulsar spin frequency is tracked using a Kalman filter concurrently with the pulse frequency modulation induced by a gravitational wave from a single source. The modulation is the sum of terms proportional to the gravitational wave strain at the Earth and at every pulsar in the array. Here we generalize previous state-space formulations of the pulsar timing array problem to include the pulsar terms; that is, we copy the pulsar terms from traditional, non-state-space analyses over to the state-space framework. The performance of the generalized Kalman filter is tested using astrophysically representative software injections in Gaussian measurement noise. It is shown that including the pulsar terms corrects for previously identified biases in the parameter estimates (especially the sky position of the source) which also arise in traditional matched-filter analyses that exclude the pulsar terms. Additionally, including the pulsar terms decreases the minimum detectable strain by $14\%$. Overall, the study verifies that the pulsar terms do not raise any special extra impediments for the state-space framework, beyond those studied in traditional analyses. The inspiral-driven evolution of the wave frequency at the Earth and at the retarded time at every pulsar in the array is also investigated.
\end{abstract}




\begingroup
\let\clearpage\relax
\endgroup
\newpage
\section{Introduction}\label{sec:intro}
Nanohertz gravitational waves (GWs) are produced by the inspiral of supermassive black hole binaries \citep[SMBHBs;][]{Rajagopal1995,Jaffe_2003, Wyithe2003,Sesana2013,McWilliams_2014,Ravi2015MNRAS.447.2772R,Burke2019, Skyes2022}. They modulate sinusoidally the times of arrival (TOAs) at the Earth of radio pulses from pulsars. A pulsar timing array \citep[PTA;][]{ Tiburzi2018, 2021hgwa.bookE...4V} simultaneously measures the TOAs from multiple pulsars to search for a coincident GW signature. Congruent evidence from PTA observations has been presented \citep{2023ApJ...951L...8A,2023arXiv230616214A,2023ApJ...951L...6R,2023RAA....23g5024X} for a stochastic GW background, which arises from the incoherent superposition of multiple SMBHB sources \citep{Allen1997,Sesana10,Christensen2019,Renzini2022}. SMBHBs that are sufficiently close and massive may be resolvable individually but no definitive detection has been claimed so far \citep{Jenet2004,Zhu2014PPTA,Babak2016,Arzoumanian2023,2023arXiv230616226A}. \newline 

The modulation in a pulsar's TOAs due to a GW has two contributions: an ``Earth term'' and ``pulsar terms'' which are proportional to the GW strain at the Earth and every pulsar respectively. Whilst the Earth term is phase coherent between all pulsars, the pulsar terms have uncorrelated phases. Consequently the pulsar terms are typically considered as a source of self-noise and dropped from many --- although not all --- standard PTA analyses \citep[e.g.][]{Sesana2010,Babak2012,Petiteau2013,Zhu2015,Taylors2016,Goldstein2018,Charisi2023arXiv230403786C}. It is acknowledged in standard analyses that dropping the pulsar terms leads to biases in the inferred parameters and reduces the detection probability, but both impacts are modest \citep{Zhupulsarterms,Chen2022,KimpsonPTA}. \newline 

\cite{KimpsonPTA} (K24 hereafter) introduced a state-space method for the detection and parameter estimation of continuous GWs from individual SMBHBs, which self-consistently tracks the intrinsic, achromatic timing noise in PTA pulsars \citep[e.g.][]{Shannon2010,Lasky2015,Caballero2016,Goncharov2021} and disentangles it from GW-induced TOA modulations. The method described in K24 complements standard PTA analyses by tracking the specific, measured, time-ordered realization of the TOA fluctuations instead of fitting for their ensemble-averaged statistics (such as their power spectral density), following the customary signal processing approach in many industrial and scientific electrical engineering applications. The method described in K24 also follows the example of many standard PTA analyses and drops the pulsar terms. In this paper we achieve two goals. First, we extend K24 to include the pulsar terms. We demonstrate how the static GW parameters and the GW phase at each pulsar can be estimated successfully by combining a Kalman filter \citep{Kalman1} with a Bayesian nested sampler \citep{Skilling, Ashton2022}, repeating the successful demonstration in K24 but with the pulsar terms now included. Second, we quantify how including the pulsar terms improves (i) the accuracy with which the static GW parameters are estimated, and (ii) the minimum detectable GW strain, compared to when the pulsar terms are omitted. We emphasize that the biases incurred by omitting the pulsar terms have been studied thoroughly in the context of standard PTA analyses \citep{Zhupulsarterms,Chen2022}; they are not new effects discovered here for the first time. Rather, the goal of this paper is to study them again in the context of the promising state-space formulation of the PTA analysis problem introduced by K24, to check whether they raise any special extra impediments beyond those studied in standard PTA analyses. The state-space formulation complements standard analysis techniques, e.g.\ based on matched filters \citep{PhysRevD.79.084030,PhysRevD.91.044048}. It does not supplant them and is likely to be most informative when run in tandem. \newline

The paper is organised as follows. In Section \ref{sec:2} we briefly review the state-space formulation of PTA data analysis introduced by K24. In Section \ref{sec:pulsar_term} we show how to include the pulsar terms via a convenient reparametrisation. In Sections \ref{sec:rep_example1} and \ref{sec:pe_and_ms} we explain how to test the updated model, inclusive of the pulsar terms, by employing synthetic data for a single representative GW source across an astrophysically relevant domain of SMBHB source parameters. In Section \ref{sec:earth_vs_psr} we quantify the parameter estimation accuracy and the minimum detectable GW strain, comparing the performance when the pulsar terms are included and excluded.. Conclusions are drawn in Section \ref{sec:discussion}. Throughout the paper we adopt natural units, with $c = G = \hbar = 1$, and the metric signature $(-,+,+,+)$. \newline

\section{State-Space Formulation}\label{sec:2}
In this section we briefly review the state-space formulation of a PTA experiment, as presented in K24. There are $N$ pulsars in the array, labelled $1\leq n\leq N$. The intrinsic spin frequency of the $n$-th pulsar, $f_{\rm p}^{(n)}(t)$, measured in the local, freely-falling rest frame of the pulsar's centre of mass, evolves according to a stochastic differential equation,  which describes secular braking combined with intrinsic, achromatic spin wandering (``timing noise''). The radio pulse frequency measured by an observer at Earth, $f_{\rm m}^{(n)}(t)$, is related to $f_{\rm p}^{(n)}(t)$ via a measurement equation, which describes the TOA modulation induced by the GW. In Section \ref{sec:psr_frequency} we define and justify the phenomenological equation of motion for $f_{\rm p}^{(n)}(t)$. In Section \ref{sec:psr_measured} we outline the measurement equation relating $f_{\rm m}^{(n)}(t)$ to $f_{\rm p}^{(n)}(t)$. In Section \ref{sec:ss_params} we summarise the static parameters of the model, including the GW source parameters, which are estimated ultimately from the data by nested sampling. The analysis method accepts as an input a sequence of pulse frequencies, instead of a sequence of pulse TOAs, in order to validate the approach in its simplest incarnation and to maintain consistency with previous work \citep{Myers2021MNRAS.502.3113M,Meyers2021,KimpsonPTA}. It will be necessary to modify the method to accept pulse TOAs when analyzing real data, a subtle extension which we defer to a forthcoming paper.

\subsection{Spin evolution} \label{sec:psr_frequency}
We assume that the rest frame spin frequency of the $n$-th pulsar evolves according to a mean-reverting Ornstein-Uhlenbeck process, described by a Langevin equation with a time-dependent drift term \citep{Vargas},
\begin{equation}
	\frac{df_{\rm p}^{(n)}}{dt} = -\gamma^{(n)}	 [f_{\rm p}^{(n)} - f_{\rm em}^{(n)} (t)] + \dot{f}_{\rm em}^{(n)}(t) +\xi^{(n)}(t) \ , 
	\label{eq:frequency_evolution}
\end{equation}
where $f_{\rm em}^{(n)}$ is the deterministic component of the evolution, an overdot denotes a derivative with respect to $t$, $\gamma^{(n)}$ is a damping constant whose reciprocal specifies the mean-reversion timescale, and $\xi^{(n)}(t)$ is a white noise stochastic process which satisfies
\begin{align}
	\langle \xi^{(n)}(t) \rangle &= 0 \ , \\
	\langle \xi^{(n)}(t) \xi^{(n')}(t') \rangle &= [\sigma^{(n)}]^2 \delta_{n,n'} \delta(t - t') \ .	\label{eq:xieqn}
\end{align}
In Equation \eqref{eq:xieqn}, $[\sigma^{(n)}]^2$ parametrizes the noise amplitude and produces characteristic root mean square fluctuations $\approx \sigma^{(n)} / [\gamma^{(n)}]^{1/2}$ in $f_{\rm p}^{(n)}(t)$ \citep{gardiner2009stochastic}. The deterministic evolution $f_{\rm em}^{(n)}(t)$ is attributed to magnetic dipole braking for the sake of definiteness, with braking index $n_{\rm em}=3$ \citep{1969ApJ...157..869G}. PTAs are typically composed of millisecond pulsars (MSPs), for which the quadratic correction due to $n_{\rm em}$ in $f_{\rm p}^{(n)}(t)$ is negligible over the observation time $T_{\rm obs} \sim 10 \, {\rm yr}$. Consequently, 	$f_{\rm em}^{(n)}(t)$ can be approximated accurately by 
\begin{equation}
	f_{\rm em}^{(n)}(t) = f_{\rm em}^{(n)}(t_1) + \dot{f}_{\rm em}^{(n)}(t_1) (t - t_1) \ , \label{eq:spinevol}
\end{equation} 
where $t_1$ labels the first TOA. \newline 

Equations \eqref{eq:frequency_evolution}--\eqref{eq:spinevol} are not unique. Rather, they offer one possible phenomenological description consistent with the main qualitative features of a typical PTA pulsar's observed spin evolution, i.e.\ random, mean-reverting, small-amplitude excursions around a smooth, secular trend \citep{NANOgrav2023,EPTA2023,Zic2023arXiv230616230Z}. A phenomenological approach is obligatory, because a predictive, first-principles theory of timing noise does not currently exist, c.f. the multiple theorized mechanisms referenced in Section 1 of K24. Langevin equations like \eqref{eq:frequency_evolution}--\eqref{eq:spinevol} have been applied successfully to analyse anomalous braking indices \citep{Vargas} and in hidden Markov model glitch searches \citep{Melatos2020ApJ...896...78M,Lower2021MNRAS.508.3251L,Dunn2022,Dunn2023MNRAS.522.5469D}. However, they are highly idealised \citep{Meyers2021,Myers2021MNRAS.502.3113M,2023MNRAS.520.2813A,Vargas}. Idealisations include (i) the exclusion of physics, which is likely to be present in reality, e.g.\ the classic, two-component, crust-superfluid structure inferred from post-glitch recoveries \citep{Baym1969,vanEysden,Alpar2017MNRAS.471.4827G,Myers2021MNRAS.502.3113M,Meyers2021}; (ii) the exclusion of non-Gaussian excursions such as L\'{e}vy flights \citep{Sornette2004}; (iii) the whiteness of $\xi^{(n)}(t)$ for MSPs in PTAs; and (iv) the formal interpretation of $d^2 f_{\rm p} / dt^2$, noting that $\xi^{(n)}(t)$ in Equation \eqref{eq:frequency_evolution} is not differentiable.

\subsection{Modulation of pulsar frequency by a GW} \label{sec:psr_measured}
In the presence of a GW, the rest-frame spin frequency of the $n$-th pulsar is related to the radio pulse frequency measured by an observer on Earth via a measurement equation,
\begin{equation}
	f_{\rm m}^{(n)}(t) = f_{\rm p}^{(n)}\left [t-d^{(n)} \right ] g^{(n)}(t) +  \varepsilon^{(n)}(t)\ ,
	\label{eq:measurement}
\end{equation}
where $d^{(n)}$ labels the distance to the $n$-th pulsar, $f_{\rm p}^{(n)}$ is evaluated at the retarded time $t-d^{(n)}$, and $\varepsilon^{(n)}(t)$ is a Gaussian measurement noise which satisfies 
\begin{align}
	\langle \varepsilon^{(n)}(t) \rangle &= 0 \ , \\
	\langle \varepsilon^{(n)}(t) \varepsilon^{(n')}(t') \rangle &= \left[\sigma_{\rm m}^{(n)}\right]^2 \delta_{n,n'} \delta(t - t') \ .	\label{eq:vareps}
\end{align}
In Equation \eqref{eq:vareps}, $[\sigma_{\rm m}^{(n)}]^2$ is the variance of the measurement noise at the telescope. In Equation \eqref{eq:measurement} the measurement function $g^{(n)}(t)$ is given by \citep[e.g.][]{Maggiore}
\begin{align}
	g^{(n)}(t) =& 1 - \frac{ H_{ij}[q^{(n)}]^i [q^{(n)}]^j }{2 [1 + \boldsymbol{n}\cdot \boldsymbol{q}^{(n)}] } \nonumber \\
	& \times \Big[\cos\left(-\Omega t +\Phi_0\right) \nonumber \\
	&- \cos \left \{-\Omega t +\Phi_0 + \Omega \left[1 + \boldsymbol{n}\cdot \boldsymbol{q}^{(n)} \right]  d^{(n)} \right \} \Big ] \ ,
	\label{eq:g_func_trig}
\end{align}
where $[q^{(n)}]^i$ labels the $i$-th coordinate component of the $n$-th pulsar's position vector $\boldsymbol{q}^{(n)}$, $\Omega$ is the constant angular frequency of the GW, $\boldsymbol{n}$ is a unit vector specifying the direction of propagation of the GW, $H_{ij}$ is the spatial part of the GW amplitude tensor, and $\Phi_0$ is the phase offset of the GW with respect to some reference time. \newline

\noindent Equation \eqref{eq:g_func_trig} assumes that (i) $\boldsymbol{q}^{(n)}$ is constant; and (ii) the GW is a monochromatic plane wave. \newline

Regarding point (i) a pulsar's sky position varies due to its own orbital motion (if it is located in a binary) as well as the rotation and revolution of the Earth. However, pulsar TOAs are defined relative to the Solar System barycentre, after correcting for the pulsar's binary motion (if any). The barycentering correction is typically applied when generating TOAs, e.g. with {\sc tempo2} \citep{tempo2,edwardstempo} and related timing software, and is inherited by $f_{\rm m}^{(n)}(t)$. Some PTA pulsars do have non-negligible proper motions of order $10^2$ km s$^{-1}$ after barycentering \citep[e.g.][]{10.1093/mnras/sty3390}, but we do not consider this effect in this introductory paper. \newline

Regarding point (ii), there are two timescales to consider. The first is the timescale set by the observation period, $T_{\rm obs}$. The second is the timescale set by the light travel time between the pulsar and the Earth, $T_{\rm light}^{(n)} = d^{(n)}/c$. Regarding the former, studies of SMBHB inspirals in the PTA context show that the GW frequency $f_{\rm gw}$ ($=\Omega / 2 \pi $) evolves over a short time-scale by an amount \citep[e.g.][]{Sesana2010}
\begin{equation}
	\Delta f_{\rm gw} = 0.05 \, \mathrm{nHz}\left(\frac{M_{\rm c}}{10^{8.5} M_{\odot}}\right)^{5 / 3}\left[\frac{f_{\rm gw}(t=t_1)}{50 \mathrm{~nHz}}\right]^{11 / 3}\left(\frac{T_{\mathrm{obs}}}{10 \mathrm{yr}}\right) \ ,
	\label{eq:f_evolution}
\end{equation}
where $M_{\rm c}$ is the chirp mass of the SMBHB, $f_{\rm gw}(t=t_1)$ is the GW frequency at the time of the first observation, and one has typically  $T_{\rm obs} \sim 10$ years. A source can be considered monochromatic if $\Delta f_{\rm gw}$ is less than the PTA frequency resolution $1/T_{\rm obs}$ \footnote{Strictly speaking, the frequency resolution is proportional to $1/T_{\rm obs}$ divided by the signal-to-noise ratio; brighter sources are resolved more accurately.}. A vast majority of SMBHBs resolved by PTAs are expected to satisfy $\Delta f_{\rm gw} < 1/T_{\rm obs}$ and we are therefore justified in treating the GW source as monochromatic over the $T_{\rm obs}$ timescale as a first approximation \citep{Sesana10,Sesana2010,Ellis2012ApJ}. \newline 

Regarding the light travel time, for SMBHBs which are sufficiently massive and have sufficiently high orbital frequencies, the source may undergo non-negligible evolution during $T_{\rm light}^{(n)}$, such that the frequency of the GW which is incident upon the Earth does not equal the frequency of the GW which is incident upon the pulsar. Most SMBHBs detectable with PTAs are not expected to satisfy this condition, but some do; for a PTA composed of pulsars with a mean distance of 1.5 kpc, 78\% of simulated SMBHBs detectable with the current IPTA undergo negligible evolution, whilst for the second phase of the Square Kilometre Array this fraction drops to 52\%; see Figure 7 in  \cite{Rosado10.1093/mnras/stv1098}. In this introductory paper, as a first pass, we focus exclusively on sources which undergo negligible evolution during $T_{\rm light}^{(n)}$, following previous investigations of pulsar-term biases in the context of standard PTA analyses \citep[e.g.][]{Zhupulsarterms,Chen2022}. However, the issue is an important one, so we perform some preliminary tests regarding how sensitively the results depend on the monochromatic assumption in Appendix \ref{sec:monochromatic}. We find that the results do not depend sensitively on the assumption that the GW source has a constant angular frequency, especially for systems with low signal-to-noise ratio (SNR). More extensive testing is deferred to future work, after the state-space method matures sufficiently (e.g.\ by tracking pulsar phase) to be applied to real, astronomical data. 

\subsection{Static parameters}\label{sec:ss_params}
The model described in Sections \ref{sec:psr_frequency} and \ref{sec:psr_measured} comprises $5N$ static parameters, that are specific to the pulsars in the array, viz.
\begin{equation}
	\boldsymbol{\theta}_{\rm psr} = \left \{ \gamma^{(n)},\sigma^{(n)}, f_{\rm em}^{(n)}(t_1),\dot{f}_{\rm em}^{(n)}(t_1),d^{(n)}\right\}_{1\leq n \leq N} \ .  \label{eq:psrparams}
\end{equation}
It also comprises seven parameters, that are specific to the GW source, viz. 
\begin{equation}
	\boldsymbol{\theta}_{\rm gw} = \left \{ h_0, \iota, \psi, \delta, \alpha, \Omega, \Phi_0 \right \} \ ,  \label{eq:params3}
\end{equation}
where $h_0$ is the characteristic wave strain, $\iota$ is the orbital inclination, $\psi$ is the polarisation angle, $\delta$ is the declination and $\alpha$ is the right ascension. These parameters enter the model through Equation \eqref{eq:g_func_trig}, with $H_{ij} = H_{ij}(h_0, \iota, \psi, \delta, \alpha)$ and $\boldsymbol{n}=\boldsymbol{n}(\delta,\alpha)$. The complete set of $7 + 5N$ static parameters is denoted by $\boldsymbol{\theta} = \boldsymbol{\theta}_{\rm gw} \cup \boldsymbol{\theta}_{\rm psr}$. While we assume no prior information about $\boldsymbol{\theta}_{\rm gw}$, there are constraints on $\boldsymbol{\theta}_{\rm psr}$ from electromagnetic observations; for example estimates of $d^{(n)}$ are accurate to $\sim$ 10$\%$ typically \citep{Cordes2002astro.ph..7156C, Verbiest2012ApJ...755...39V, Desvignes2016,Yao2017}.

\section{Inference with the pulsar terms}\label{sec:pulsar_term}
K24 developed a likelihood-based Bayesian framework to infer the static parameters in Section \ref{sec:ss_params} and select between models with and without a GW. Given a temporal sequence of noisy measurements $\boldsymbol{Y}(t)$, the posterior distribution of $\boldsymbol{\theta}$ is calculated by Bayes' Rule,
\begin{equation}
	p(\boldsymbol{\theta} | \boldsymbol{Y}) = \frac{\mathcal{L}(\boldsymbol{Y} | \boldsymbol{\theta}) \pi(\boldsymbol{\theta})}{\mathcal{Z}} \ , \label{eq:posterior_distrib}
\end{equation}
where $\mathcal{L}(\boldsymbol{Y}| \boldsymbol{\theta})$ is the likelihood calculated with a Kalman filter \citep{Kalman1}, as discussed in Appendix \ref{sec:kalman}, $\pi(\boldsymbol{\theta})$ is the prior distribution of $\boldsymbol{\theta}$, and $\mathcal{Z}$ is the marginalised likelihood or evidence,
\begin{equation}
	\mathcal{Z} = \int d \boldsymbol{\theta} \mathcal{L}(\boldsymbol{Y} | \boldsymbol{\theta})  \pi(\boldsymbol{\theta})  \ . \label{eq:model_evidence}
\end{equation}
The measurements are ${\boldsymbol{Y}} = \{ f^{(n)}_{\rm m}(t_1), \dots, f^{(n)}_{\rm m}(T_{\rm obs})  \}_{1 \leq n \leq N}$. Given a specific realisation of $\boldsymbol{Y}$, $\mathcal{L}(\boldsymbol{Y}| \boldsymbol{\theta})$ is a function of $\boldsymbol{\theta}$. Appendix \ref{sec:kalman} explains how to compute ${\mathcal{L}}({\boldsymbol{Y}}|{\boldsymbol{\theta}})$ from $f_{\rm m}^{(n)}(t)$ by discretizing the dynamical equations \eqref{eq:frequency_evolution}--\eqref{eq:spinevol} and the measurement equations \eqref{eq:measurement}--\eqref{eq:g_func_trig} and solving them recursively using a Kalman filter. Nested sampling \citep{Skilling} is used to estimate $p(\boldsymbol{\theta} | \boldsymbol{Y})$ and $\mathcal{Z}$ given ${\mathcal{L}}({\boldsymbol{Y}}|{\boldsymbol{\theta}})$. Appendix \ref{sec:nested_sampling} reviews nested sampling briefly as applied in this paper. Appendix \ref{sec:workflow} summarizes the workflow of the combined Kalman filter and nested sampler. Appendices \ref{sec:kalman}--\ref{sec:workflow} are designed to equip the interested reader to reproduce the key results in Sections \ref{sec:pe_and_ms} and \ref{sec:earth_vs_psr}. \newline

The inference procedure in K24 drops the pulsar term on the last line of Equation \eqref{eq:g_func_trig}, in keeping with several other PTA analyses \citep[e.g.][]{Sesana2010,Babak2012,Petiteau2013,Zhu2015,Taylors2016,Goldstein2018,Charisi2023arXiv230403786C}. In this section we show how to remove this limitation, following previous authors \citep{Zhupulsarterms,Chen2022}. In Section \ref{sec:earth_term} we review briefly how the pulsar terms are dropped from typical PTA analyses as well as K24. In Section \ref{sec:pulsar_term2} we introduce a convenient reparametrisation of the pulsar terms for likelihood-based inference methods such as nested sampling. The new parametrisation is validated with synthetic data in Section \ref{sec:rep_example1}.

\subsection{Earth term}\label{sec:earth_term}
The GW modulates radio pulses according to Equation \eqref{eq:g_func_trig}. The modulation is composed of an Earth term, proportional to $\cos(-\Omega t + \Phi_0)$, and a pulsar term, proportional to $\cos \left \{-\Omega t +\Phi_0 + \Omega \left[1 + \boldsymbol{n}\cdot \boldsymbol{q}^{(n)} \right]  d^{(n)} \right \}$. The Earth term describes the GW phase at the observer on Earth, while the pulsar term describes the GW phase at the $n$-th pulsar. The Earth term depends only on the GW source parameters and is common across all pulsars. In contrast the pulsar term is a function of $d^{(n)}$ and $\boldsymbol{q}^{(n)}$ and varies between pulsars. \newline

The phases of the pulsar terms are related unpredictably, which is why some published PTA analyses approximate the pulsar terms collectively as a source of self-noise. The pulsar term depends on $d^{(n)}$ which is generally poorly constrained by electromagnetic observations, with uncertainties greater than the typical GW wavelength. Consequently the pulsar term is often --- although not always --- dropped in standard PTA analyses \citep[e.g.][]{Sesana2010,Babak2012,Petiteau2013,Zhu2015,Taylors2016,Goldstein2018,Charisi2023arXiv230403786C}. Dropping the pulsar term is also convenient computationally because it reduces the dimensionality of the parameter space; the $N$ values of $d^{(n)}$ are not inferred. Within the state-space model described in Section \ref{sec:2}, dropping the pulsar terms equates to using a modified measurement equation,
\begin{equation}
	f_{\rm m}^{(n)}(t) = f_{\rm p}^{(n)}\left [t-d^{(n)} \right ] g^{(n)}_{\rm Earth}(t) + \varepsilon^{(n)}(t) \ , 
	\label{eq:measuremen_earth}
\end{equation}
with
\begin{equation}
	g^{(n)}_{\rm Earth}(t) = 1 - \frac{ H_{ij}[q^{(n)}]^i [q^{(n)}]^j}{2[1 + \boldsymbol{n}\cdot \boldsymbol{q}^{(n)}] }  \cos(-\Omega t +\Phi_0)  \ .
	\label{eq:g_func_trig_earth}
\end{equation}
However, dropping the pulsar term leads to well-known parameter estimation biases, especially in $\alpha$ and $\delta$, and reduces the detection probability by $\sim 5 \%$ \citep{Zhupulsarterms,Chen2022,KimpsonPTA}. 

\subsection{Reparametrisation of the pulsar terms} \label{sec:pulsar_term2}
In this paper, we generalize the Kalman filter analysis in K24 to include the pulsar terms. To do so, we define a new parameter 
\begin{equation}
	\chi^{(n)} = \left\{ \Omega \left[ 1 + \boldsymbol{n}\cdot \boldsymbol{q}^{(n)} \right]  d^{(n)} \right \} \mod 2 \pi \ , \label{eq:chi_param}
\end{equation}
such that Equation \eqref{eq:g_func_trig} becomes
\begin{align}
	g^{(n)}(t) =& 1 - \frac{ H_{ij}[q^{(n)}]^i [q^{(n)}]^j }{2 [1 + \boldsymbol{n}\cdot \boldsymbol{q}^{(n)}] } \nonumber \\
	& \times \Big \{\cos\left(-\Omega t +\Phi_0\right) \nonumber \\
	&- \cos \left [-\Omega t +\Phi_0 + \chi^{(n)} \right ] \Big \} \ .
	\label{eq:g_func_trig_chi}
\end{align}
The reparametrisation in terms of $\chi^{(n)}$ treats the pulsar-dependent phase correction $\Omega \left[ 1 + \boldsymbol{n}\cdot \boldsymbol{q}^{(n)} \right]  d^{(n)}$ in the argument of the cosine of the pulsar term as a composite parameter to be inferred for each pulsar. In principle, it is possible to disentangle $\Omega$, $\boldsymbol{n}$, and $d^{(n)}$ and infer them individually, if $N$ is large enough; for example, $\boldsymbol{n}$ appears independently in the phase of the pulsar term and in the denominator of the first line of Equation \eqref{eq:g_func_trig}. In practice, however, the likelihood surface is highly corrugated along the $d^{(n)}$-axis, and the prior on $d^{(n)}$ is broad compared to the wavelength $2\pi/\Omega$, so $d^{(n)}$ and hence $\Omega$ are not identifiable for reasonable $N \lesssim 10^2$. By replacing $d^{(n)}$ with $\chi^{(n)}$, we obtain a smooth likelihood function, whose maximum is located efficiently and accurately by the nested sampler. This point is discussed in more detail in Appendix \ref{sec:psr_term_challenges}.\newline 
 
The new parametrisation does not increase the dimension of the parameter space, because we trade $d^{(n)}$ for $\chi^{(n)}$. Once $\Omega$, ${\boldsymbol{n}}$, and $\chi^{(n)}$ are inferred, it is possible to solve for $d^{(n)}$, albeit not uniquely; $d^{(n)}$ can be inferred up to an integer multiple of the Doppler-shifted wavelength, due to the ${\rm mod} \, 2 \pi$ operation in Equation \eqref{eq:chi_param}. However, this ambiguity is unavoidable, whether we replace $d^{(n)}$ with $\chi^{(n)}$ or not, and occurs in every PTA analysis. The static parameters specific to the pulsars in the array (c.f. Equation \eqref{eq:psrparams}) are now
\begin{equation}
	\boldsymbol{\theta'}_{\rm psr} = \left \{ \gamma^{(n)},\sigma^{(n)}, f_{\rm em}^{(n)}(t_1),\dot{f}_{\rm em}^{(n)}(t_1),\chi^{(n)} \right\}_{1\leq n \leq N} \ ,  \label{eq:psrparamsnew}
\end{equation}
whilst $\boldsymbol{\theta}_{\rm gw}$  remains unchanged. We define the complete set of $7 +5N$ static parameters to be inferred as $\boldsymbol{\theta'} = \boldsymbol{\theta}_{\rm gw} \cup \boldsymbol{\theta'}_{\rm psr}$.

\section{Validation with synthetic data}\label{sec:rep_example1}
In the rest of this paper, we test the performance of the PTA analysis scheme in K24, once it is generalized to include the pulsar terms in the inference model as described in Section 3. We refer the reader to the appendices for detailed instructions about how to implement the Kalman filter (Appendix \ref{sec:kalman}) and nested sampler (Appendix \ref{sec:nested_sampling}) and integrate them in a unified workflow (Appendix \ref{sec:workflow}). In this section, by way of preparation, we explain how to generate the synthetic data employed in the tests, namely noisy frequency time series $f_{\rm m}^{(n)}(t)$ for $1 \leq n \leq N$. Tests are performed for multiple random realizations of $\xi^{(n)}(t)$ in order to quantify the irreducible ``cosmic'' variance in the inference output (e.g.\ estimates of ${\boldsymbol{\theta}}_{\rm gw}$). Every real PTA analysis witnesses a unique realization of $\xi^{(n)}(t)$ --- the actual, astronomical one --- but there is no way to determine where this realization lies within the admissible statistical ensemble. \newline 

\begin{table*}
	\centering
		\begin{tabular}{lccll}
			\toprule
			Set&Parameter & Injected value & Units & Prior  \\
			\hline
			\multirow{7}{2mm}{$\boldsymbol{\theta}_{\rm gw}$} & $\Omega$       & $5 \times 10^{-7}$ & Hz & LogUniform($10^{-9}$, $10^{-5}$) \\
			& $\alpha$          & $1.0$  & rad & Uniform($0, 2 \pi $)\\
			& $\delta$              & $1.0$  & rad & Cosine($-\pi/2, \pi/2$) \\
			& $\psi$              & $2.50$ & rad & Uniform($0, 2 \pi $) \\
			& $\Phi_0$          & $0.20$ & rad & Uniform($0, 2 \pi $) \\
			& $h_0$            & $5 \times 10^{-15}$ & --- & LogUniform($10^{-15}$, $10^{-9}$) \\
			& $\iota$             & $1.0$ & rad & Sin($0, \pi$) \\ 
			\hline
			\vspace{1mm}& $f_{\rm em}^{(n)} (t_1)$       & $f_{\rm ATNF}^{(n)}$ & Hz & Uniform$\left[f_{\rm ATNF}^{(n)} - 10^3 \eta^{(n)}_{f}, f_{\rm ATNF}^{(n)} + 10^3 \eta^{(n)}_{f} \right]$ \\
      \multirow{2}{2mm}{$\boldsymbol{\theta'}_{\rm psr}$} & $\dot{f}_{\rm em}^{(n)} (t_1)$       & $\dot{f}_{\rm ATNF}^{(n)}$ & s$^{-2}$ & Uniform$\left[ \dot{f}_{\rm ATNF}^{(n)} - 10^3 \eta^{(n)}_{\dot{f}}, \dot{f}_{\rm ATNF}^{(n)} + 10^3 \eta^{(n)}_{\dot{f}} \right]$ \\
		     & $\sigma^{(n)}$              & $\sigma_{\rm SC}^{(n)}$ & $s^{-3/2}$ & LogUniform$ \left [10^{-2} \sigma_{\rm SC}^{(n)}, 10^2 \sigma_{\rm SC}^{(n)} \right ]$ \\
			& $\gamma^{(n)}$              & $10^{-13}$ & s$^{-1}$ & --- \\
			\vspace{1mm} &  $\chi^{(n)}$       &$\Omega \left[ 1 + \boldsymbol{n}\cdot \boldsymbol{q}^{(n)}_{\rm ATNF} \right]  d_{\rm ATNF}^{(n)} $  & rad & Uniform($0, 2 \pi $) \\
			\bottomrule
		\end{tabular}
		\caption{Injected static parameters used to generate synthetic data to validate the analysis scheme including the pulsar terms in Equation \eqref{eq:g_func_trig_chi}. The prior used for Bayesian inference is also displayed (rightmost column).  The top and bottom sections of the table contain $\boldsymbol{\theta}_{\rm gw}$ and $\boldsymbol{\theta'}_{\rm psr}$ respectively. The subscript ``ATNF'' denotes values obtained from the ATNF pulsar catalogue as described in Section \ref{sec:rep_example1}. The subscript ``SC'' on $\sigma^{(n)}$ indicates that the injected value is calculated from Equation \eqref{eq:sigmap_f} and the empirical timing noise model for MSPs in \protect \cite{Shannon2010}. The quantities $\eta^{(n)}_{f}$ and $\eta^{(n)}_{\dot{f}}$ are the uncertainties in $f^{(n)}_{\rm em} (t_1)$ and $\dot{f}^{(n)}_{\rm em} (t_1)$ respectively, as quoted in the ATNF catalogue. We do not infer $\gamma^{(n)} \sim 10^{-5} T_{\rm obs}$ for simplicity, so no prior is set. The priors on $\boldsymbol{\theta'}_{\rm psr}$ are justified in Appendix \ref{sec:set_priors}.
		}
		\label{tab:parameters_and_priors}
	\end{table*}
In order to synthesize $\boldsymbol{Y} = \{f_{\rm m}^{(1)}(t), \dots, f_{\rm m}^{(N)}(t) \}$, we integrate Equations \eqref{eq:frequency_evolution}--\eqref{eq:spinevol} numerically using a Runge-Kutta It$\hat{\text{o}}$ integrator implemented in the \texttt{sdeint} python package \footnote{\url{https://github.com/mattja/sdeint}}. This produces random realizations of $f_{\rm p}^{(n)}(t)$ for $1\leq n \leq N$, which we convert to $f_{\rm m}^{(n)}(t)$ via Equations \eqref{eq:measurement}--\eqref{eq:g_func_trig}. The numerical solutions depend on how we choose $\boldsymbol{\theta}_{\rm psr}$, ${\boldsymbol{q}}^{(n)}$ and $\sigma_{\rm m}$ or, equivalently, how we specify the configuration of a synthetic PTA. In Section  \ref{sec:pe_and_ms} we describe how we choose the remaining elements of $\boldsymbol{\theta}$, namely $\boldsymbol{\theta}_{\rm gw}$. This latter step is equivalent to specifying the synthetic SMBHB source and differs from one test to the next according to the goal of the test.  \newline

In this paper we adopt for consistency the same $\boldsymbol{\theta}_{\rm psr}$ values as in K24, i.e. the $N=47$ MSPs in the 12.5-year NANOGrav dataset \citep{2020ApJ...905L..34A}. We assume all pulsars are observed with cadence $T_{\rm cad} = 1 \,{\rm week}$ over a 10 year period. Fiducial values for ${\boldsymbol{q}}^{(n)}$, $d^{(n)}$, $f_{\rm em}^{(n)}(t_1)$, and $\dot{f}^{(n)}_{\rm em}(t_1)$ are read from the Australia Telescope National Facility (ATNF) pulsar catalogue \citep{Manchester2005} using the \texttt{psrqpy} package \citep{psrqpy}. No direct measurements exist for $\gamma^{(n)}$. The mean reversion timescale typically satisfies $[\gamma^{(n)}]^{-1} \gg T_{\rm obs}$ \citep{Price2012,Myers2021MNRAS.502.3113M,Meyers2021,Vargas}; in this paper, for the sake of simplicity, we fix $\gamma^{(n)} = 10^{-13}$ s$^{-1}$ for all $n$. No direct measurements exist for $\sigma^{(n)}$ either. We relate $\sigma^{(n)}$ to the root mean square TOA noise $\sigma^{(n)}_{\rm TOA}$ accumulated over an interval of length $T_{\rm cad}$ by
\begin{eqnarray}
	\sigma^{(n)} \approx \sigma_{\rm TOA}^{(n)} f_{\rm p}^{(n)}(t_1) {T_{\rm cad}}^{-3/2} \ . \label{eq:sigmap_f}
\end{eqnarray}
As in K24, the empirical timing noise model for MSPs from \cite{Shannon2010ApJ...725.1607S}, applied to the 12.5-year NANOGrav dataset, implies $\text{median} [\sigma^{(n)}] = 5.51 \times 10^{-24} $ s$^{-3/2}$, $\min [ \sigma^{(n)} ] = 1.67 \times 10^{-26}$s$^{-3/2}$ for PSR J0645+5158 and $\max [ \sigma^{(n)} ] = 2.56 \times 10^{-19}$ s$^{-3/2}$ for PSR J1939+2134. \newline 
 
In a similar vein, $\sigma_{\rm m}^{(n)}$ can be related to, $\sigma_{\rm TOA}^{(n)}$, by 
\begin{equation}
	\sigma_{\rm m}^{(n)} \approx f_{\rm p}^{(n)}(t_1) \sigma_{\rm TOA}^{(n)} \ {T_{\rm cad}}^{-1} \ . \label{eq:sigma_m_eqn}
\end{equation}
For an MSP with $f_{\rm p}^{(n)} \sim 0.1$ kHz, $T_{\rm cad} = 1 \, {\rm week}$, and $\sigma_{\rm TOA}^{(n)} \sim 1 \mu$s,  Equation \eqref{eq:sigma_m_eqn} implies $\sigma_{\rm m}^{(n)} \sim 10^{-10}$ Hz. The most accurately timed pulsars have $\sigma_{\rm TOA}^{(n)} \sim 10 $ ns and $\sigma_{\rm m}^{(n)} \sim 10^{-12}$ Hz. In this paper, for simplicity and the sake of definiteness, we fix $\sigma_{\rm m}^{(n)} = 10^{-11}$ Hz for all $n$, and take it as known \textit{a priori} rather than a parameter to be inferred. When analysing real data this assumption is easily relaxed. Although $\sigma_{\rm m}^{(n)}$ is assumed to be the same for every pulsar, $f_{\rm m} ^{(n)}$ is constructed from a different random realisation of $\varepsilon^{(n)}(t)$ for each pulsar. \newline

\section{Parameter estimation} \label{sec:pe_and_ms} 
In this section, we apply the Kalman filter and nested sampler to calculate the joint posterior probability distribution $p({\boldsymbol{\theta'}} | {\boldsymbol{Y}})$ and compare it to the known, injected values. The procedure is undertaken for multiple realisations of $\xi^{(n)}(t)$ and $\varepsilon^{(n)}(t)$, and we estimate $\boldsymbol{\theta'}$ independently for each realisation. The aims are (i) to demonstrate that the analysis scheme works, i.e.\ that it converges to a well-behaved, unimodal posterior for multiple noise realisations; (ii) to give a preliminary sense of its accuracy; and (iii) to quantify the natural random dispersion in the one-dimensional posterior medians. Quantifying the dispersion is important since it is a practical measure of the scheme's accuracy when it is applied to real astronomical data, where the true parameter values and specific noise realisations are unknown (i.e. cosmic variance). The injected $\boldsymbol{\theta}_{\rm gw}$ values are selected to be astrophysically representative, as per the top section of Table \ref{tab:parameters_and_priors}. The static pulsar parameters $\boldsymbol{\theta'}_{\rm psr}$ are specified in Section \ref{sec:rep_example1}. All injected static parameters $\boldsymbol{\theta}'$ are summarised in the second column of Table \ref{tab:parameters_and_priors}. The specification of the priors on $\boldsymbol{\theta'}$ is described in Appendix \ref{sec:set_priors} and summarised in the rightmost column of Table \ref{tab:parameters_and_priors}. \newline 

In Section \ref{sec:rep_smbh_source} we apply the analysis scheme to synthetic data and start by estimating $\boldsymbol{\theta}_{\rm gw}$ for a  particular, arbitrary, representative choice of $\boldsymbol{\theta}_{\rm gw}$, i.e. the SMBHB system. The scheme is validated on multiple noise realisations of the pulsar process noise $\xi^{(n)}(t)$ and the detector measurement noise $\varepsilon^{(n)}(t)$ in order to test the scheme multiple times and quantify the  irreducible cosmic variance in the parameter estimates. In Section \ref{sec:chi_estim} we inspect and verify the estimates of  $\chi^{(n)}$. In Section \ref{sec:timing_parameters} we briefly discuss the estimates of the remaining $4N$ parameters in $\boldsymbol{\theta'}_{\rm psr}$. In Section \ref{sec:parameter_space} we extend the tests across a broader parameter domain and consider a range of astrophysically relevant SMBHB source parameters $\boldsymbol{\theta}_{\rm gw}$.

\subsection{Representative SMBHB source}\label{sec:rep_smbh_source}

Figure \ref{fig:corner_plot_1} displays the posterior distribution of $\boldsymbol{\theta}_{\rm gw}$ for ten noise realisations in the form of a traditional corner plot for the representative SMBHB source parametrized in the top portion of Table \ref{tab:parameters_and_priors}. The histograms are one-dimensional posteriors, marginalized over the six other parameters. Each coloured curve corresponds to a different noise realisation. The solid orange line marks the known injected value. The two-dimensional contours mark the (0.5, 1, 1.5, 2)-sigma level surfaces. All histograms and contours are consistent with a unimodal joint posterior, which peaks near the known, injected values. There is scant evidence of railing against the prior bounds. There is no strong evidence for correlations between parameter pairs, e.g.\ banana-shaped contours, although weak correlations are evident between $\Omega$ and $\Phi_0$, $\psi$ and $\alpha$, and $\iota$ and $h_0$. The injected value falls within the 90\% credible interval of the one-dimensional marginalized posteriors in 60 out of the 70 possible combinations of the seven parameters and 10 realizations. Indeed, the injected value falls within the 90\% credible interval for $\Omega, \Phi_0, \psi$, and  $\iota$ 40 out of 40 times. \newline  

There is appreciable natural dispersion in the one-dimensional posterior medians between realisations. We quantify the degree of dispersion using the coefficient of variation,
\begin{eqnarray}
	CV = \mu_{\rm med}/\sigma_{\rm med}
\end{eqnarray}
 where $\mu_{\rm med}$ is the mean of the 10 posterior medians and $\sigma_{\rm med}$ is their standard deviation. The minimum and maximum $CV$ are 0.2\% and 52\% for $\Omega$ and $\Phi_0$ respectively. The remaining parameters typically satisfy $CV$ $\lesssim 10 \%$. Analogous results for synthetic data with higher strain ($h_0 = 1 \times 10^{-12}$) are presented in Appendix \ref{sec:app_high_SNR} (see Figure \ref{fig:corner_high_snr_appendix}) for completeness. The dispersion between noise realisations decreases as $h_0$ increases, and the one-dimensional posteriors for $\iota$ and $h_0$ are more symmetric about the injected value. \newline 

\subsection{GW phase at each pulsar} \label{sec:chi_estim}
\begin{figure*}
	\includegraphics[width=\textwidth, height =\textwidth ]{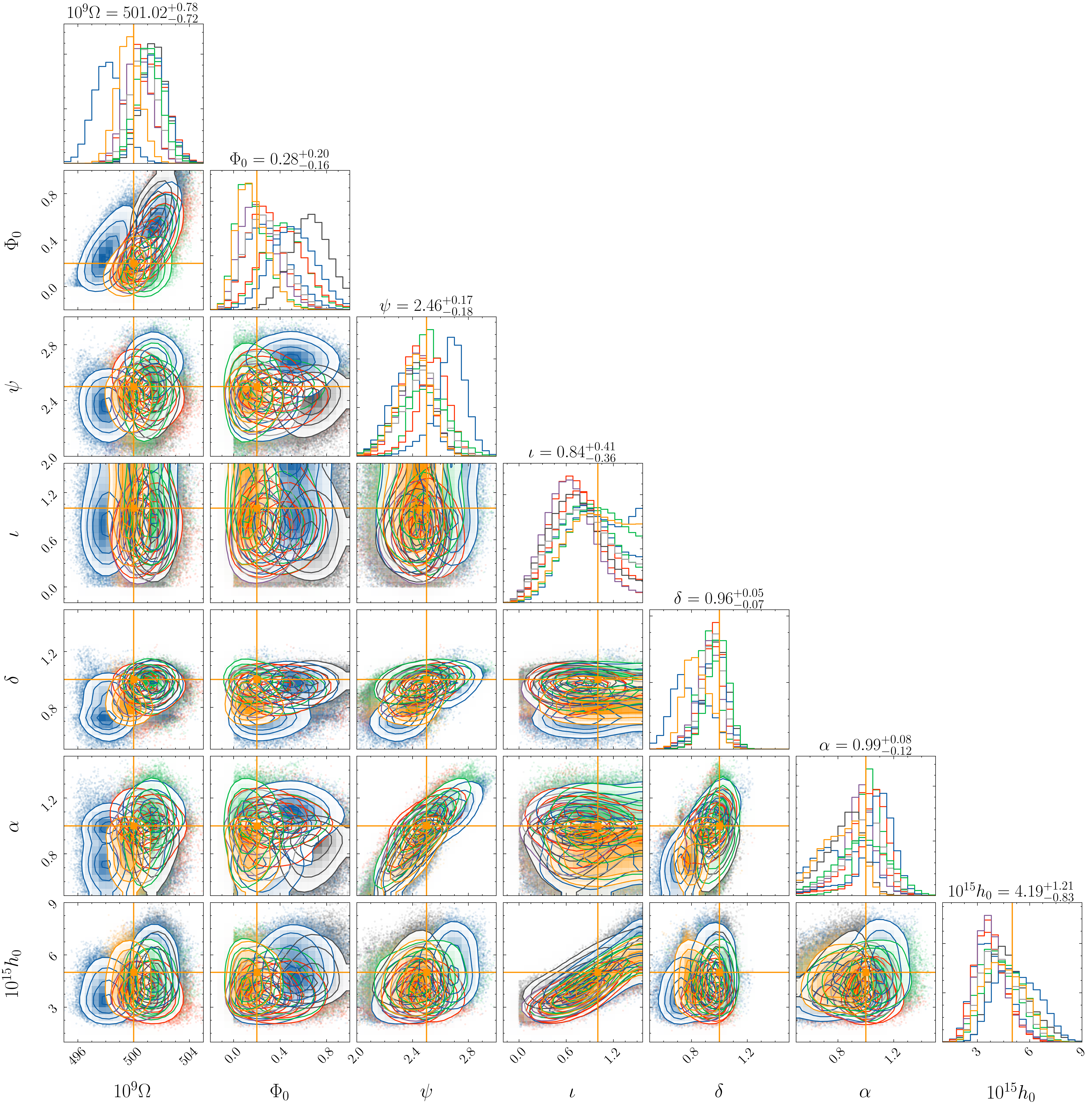}
	\caption[Caption of LOF]{Posterior distribution of the GW source parameters $\boldsymbol{\theta}_{\rm gw}$ for the representative system in Table \ref{tab:parameters_and_priors}, for 10 realisations of the noise processes, with curves coloured uniquely per realisation. The horizontal and vertical orange lines indicate the true injected values. The contours in the two-dimensional histograms mark the (0.5, 1, 1.5, 2)-$\sigma$ levels after marginalizing over all but two parameters. The one-dimensional histograms correspond to the joint posterior distribution marginalized over all but one parameter. The supertitles of the one-dimensional histograms record the median and the 0.16 and 0.84 quantiles of the median realisation. In this context the median realisation is defined as follows: a posterior is generated for all 10 realisations; the medians of the posteriors are ranked in ascending order; the median of the ranked list is associated with the median realisation. We plot the scaled variables $10^9 \Omega$ (units: ${\rm rad \, s^{-1}}$) and $10^{15} h_0$. The Kalman filter and nested sampler estimate accurately all seven parameters in ${\boldsymbol{\theta}}_{\rm gw}$. The horizontal axes span a subset of the prior domain for all seven parameters. The known, injected value lies within the 90\% credible interval for 60 out of the 70 combinations of seven parameters and ten noise realizations. There is appreciable dispersion among the peaks of the one-dimensional posteriors: the maximum coefficient of variation is $CV = 52 \%$ for $\Phi_0$, and the minimum is $CV = 0.2 \%$ for $\Omega$.}
	\label{fig:corner_plot_1}
\end{figure*}

\begin{figure}
	\includegraphics[width=\columnwidth, height =\columnwidth ]{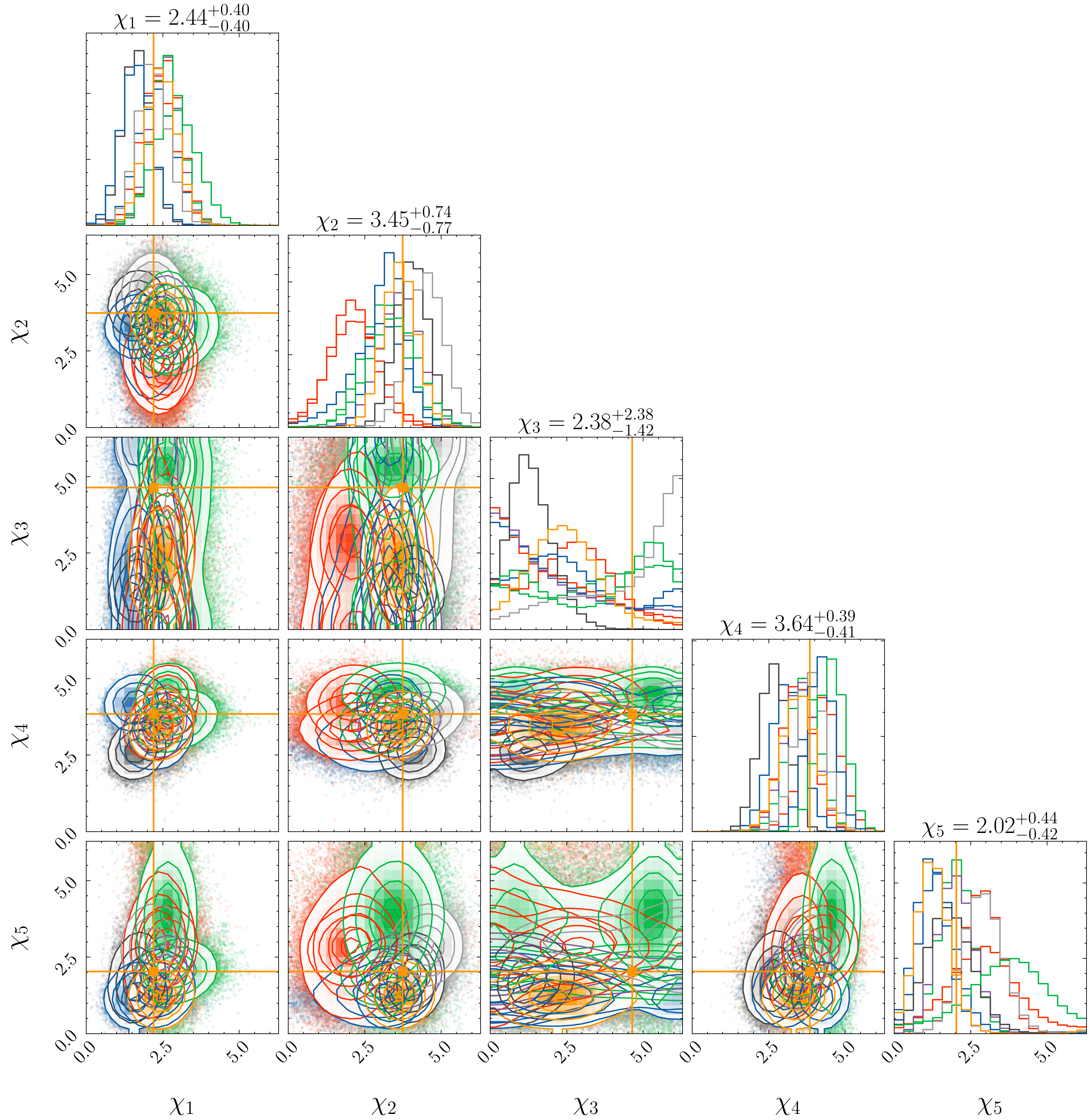}
	\caption{Same as Figure \ref{fig:corner_plot_1}	but for the static parameters $\chi^{(1)}, \dots, \chi^{(5)}$ (the remaining pulsar phases $\chi^{(6)}, \dots, \chi^{(N)}$ are omitted for readability). Consistent unimodal posteriors are obtained across the ten noise realisations for four out of the five displayed parameters. No consistent posteriors are obtained for $\chi^{(3)}$ (third row, third column), because this pulsar's sky position is close to the synthetic GW source. The known, injected value lies within the 90\% credible interval for 38 out of the 50 combinations of five parameters and ten noise realizations.}
	\label{fig:corner_plot_3}
\end{figure}

It is important to check how well the pulsar-term phase $\chi^{(n)}$ is estimated, since the reparameterization involving $\chi^{(n)}$ is a key feature of our approach, as explained in Section \ref{sec:pulsar_term2}. In short, it turns out that $\chi^{(n)}$ is estimated accurately for all $n$, except under special circumstances. Figure \ref{fig:corner_plot_3} displays the results as a corner plot for the representative subset $\chi^{(1)}, \dots, \chi^{(5)}$ over the ten noise realisations, in the same style as Figure \ref{fig:corner_plot_1}. We do not display the corner plot for $\chi^{(6)}$, $\dots$, $\chi^{(47)}$ because it is too big. With the exception of the results for $\chi^{(2)}$ (discussed below), all histograms and contours are consistent with a unimodal joint posterior, which peaks near the known, injected values. There is no evidence for correlations between parameter pairs. Most phases are recovered unambiguously across all noise realisations; for the 47 $\chi^{(n)}$ parameters (i.e. not just the five plotted in Figure \ref{fig:corner_plot_3}), the injected value is  contained within the 90\% credible interval of the one-dimensional marginalized posteriors in 383 out of the 470 possible combinations of the 47 parameters and ten realizations. The next paragraph explains why 383 out of 470 is fewer than 90\%. \newline 

Sometimes, albeit rarely, $\chi^{(n)}$ is not estimated consistently across multiple realizations for some $n$, e.g $\chi^{(3)}$ (third row, third column of Figure \ref{fig:corner_plot_3}) corresponding to PSR J0340+4130. Out of all the pulsars in the array, this object matches most closely the direction to the synthetic GW source, with $\boldsymbol{n} \cdot \boldsymbol{q}^{(2)} = -0.96$ and $\cos \chi^{(2)} \approx 0.07$. The ability to accurately infer $\chi^{(2)}$ improves with the SNR. For example, for $h_0 = 1 \times 10^{-12}$, $\chi^{(2)}$ is recovered unambiguously; see Figure \ref{fig:corner_high_snr_appendix_chi} in Appendix \ref{sec:app_high_SNR}. \newline

\subsection{Timing parameters} \label{sec:timing_parameters}
Similar results are obtained for the 4$N$ parameters in $\boldsymbol{\theta}_{\rm psr}'$ besides $\chi^{(n)}$. The injected values are recovered unambiguously across the 10 noise realisations. For the sake of brevity we do not display the calculated posterior distributions, because inferring $\boldsymbol{\theta}_{\rm gw}$ is the main focus of this paper and most published PTA analyses. In short, the estimates of $f_{\rm em}(t_1)$ and $\dot{f}_{\rm em}(t_1)$ are guided into narrow ranges by the narrow priors. The one-dimensional posteriors inferred for $\sigma^{(n)}$ are generally broader due to the broader prior, but contain the injection within the  90\% credible interval in a majority of cases. We remind the reader that $\gamma^{(n)}$ is not estimated in this paper; typically it satisfies $\gamma^{(n)} \sim 10^{-5} T_{\rm obs}$ \citep{Price2012,Myers2021MNRAS.502.3113M,Meyers2021,Vargas}, so its influence is muted. \newline

\subsection{Exploring the SMBHB parameter domain} \label{sec:parameter_space}

\begin{figure}
	\centering
	\subfloat[a][]{\includegraphics[width=\columnwidth]{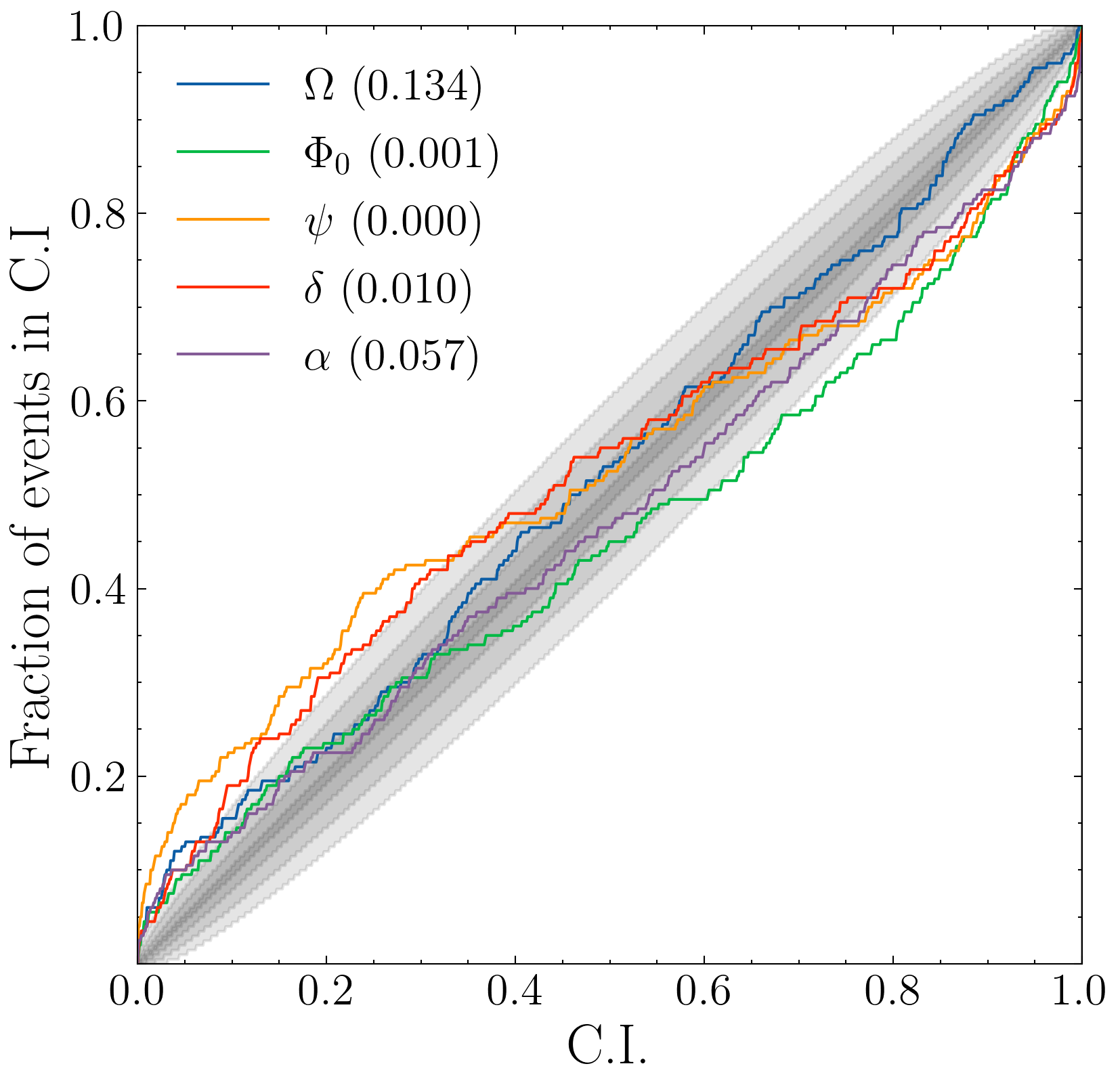} \label{fig:parameter_space}} \\
	\subfloat[b][]{\includegraphics[width=\columnwidth]{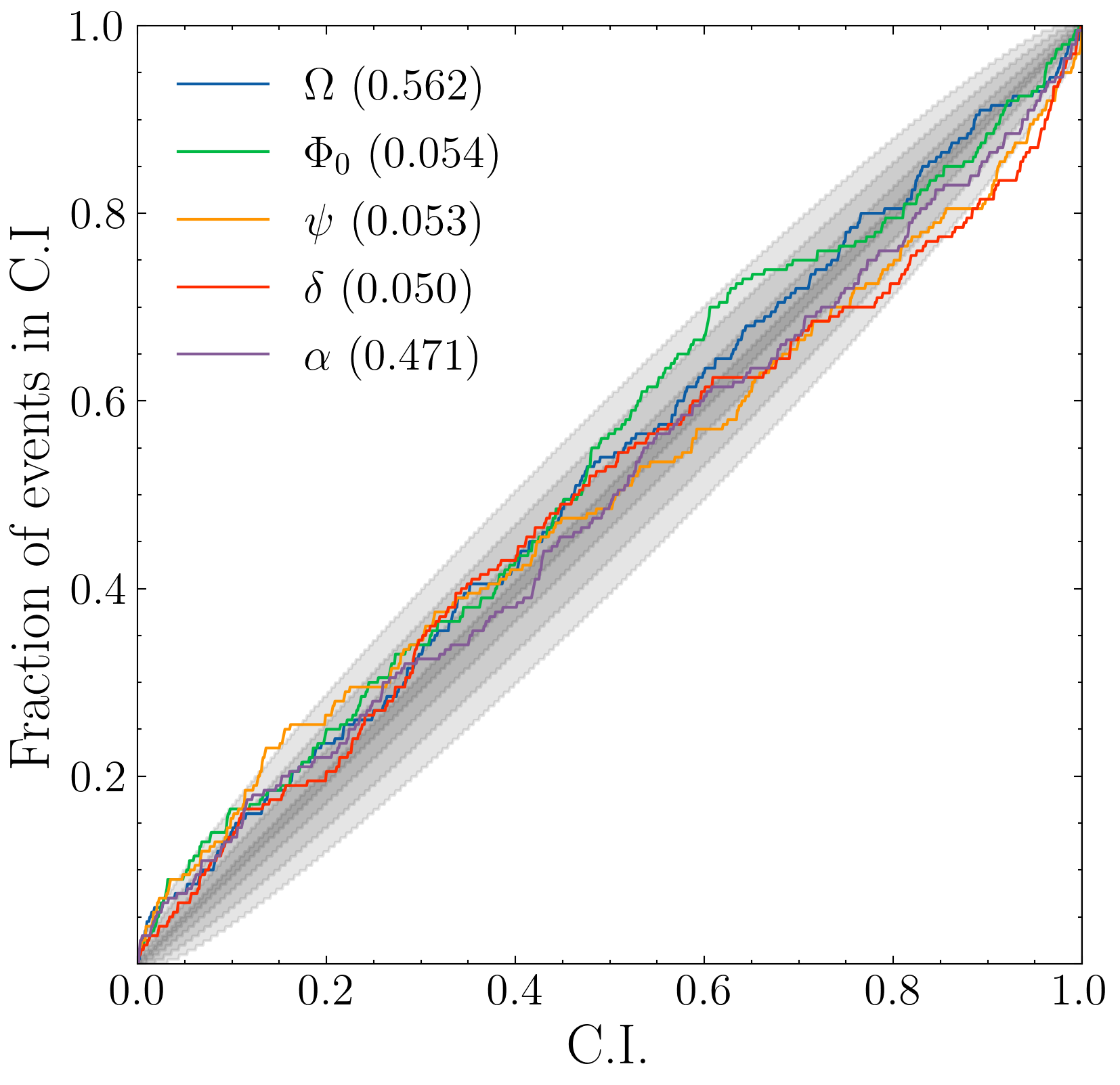} \label{fig:parameter_space2}}
	\caption{Accuracy of SMBHB parameter estimation across an astrophysically plausible domain. (a) Fraction of injections included within a given credible interval of the estimated posterior, as a function of the credible interval itself (i.e. PP plot). The injections are 200 simulated GW sources generated by drawing randomly five parameters in $\boldsymbol{\theta}_{\rm gw}$ from the prior distributions in Table \ref{tab:parameters_and_priors}. Each coloured curve corresponds to a different parameter (see legend). The parameters $h_0$ and $\iota$ are fixed at $5 \times 10^{-15}$ and 1.0 rad respectively in order to maintain an approximately constant SNR. The grey shaded contours label the $1\sigma$, $2\sigma$ and 3$\sigma$ confidence intervals. For parameters with well estimated posteriors, the PP curve should fall along the diagonal of unit slope. $\Omega$ and $\alpha$ are generally well-estimated (i.e. the curves lie close to the unit diagonal). The remaining three parameters, $\psi$, $\phi_0$ and $\delta$ show modest evidence of being over-constrained and stray outside the shaded region. (b) Same as (a) but now assuming that $\chi^{(1)}, \dots, \chi^{(47)}$ are known exactly \textit{a priori} (i.e. with a delta-function prior). The excursions from the shaded region seen in Figure  \ref{fig:parameter_space} are reduced; nearly all parameters, with the exception of $\delta$, lie wholly within the shaded region. $p$-values
	are shown for each of the parameters.} \label{fig:AB}
\end{figure}

Section \ref{sec:rep_smbh_source} focuses on a single representative source, summarised in Table \ref{tab:parameters_and_priors}. In this section we test the method for a range of sources, varying $\boldsymbol{\theta}_{\rm gw}$. The aim is to verify that the analysis scheme works across an astrophysically relevant domain and that the arbitrary choice of $\boldsymbol{\theta}_{\rm gw}$ in Table \ref{tab:parameters_and_priors} is not advantageous by accident. \newline 

We analyse 200 injections constructed by fixing $h_0 = 5 \times 10^{-15}$ and $\iota =1.0$ rad and drawing the remaining five elements of $\boldsymbol{\theta}_{\rm gw}$ randomly from the prior distributions defined in Table \ref{tab:parameters_and_priors}. We fix $h_0$ and $\iota$ in order to maintain an approximately constant SNR across the 200 injections. For each injection we compute the posterior distribution of ${\boldsymbol{\theta}}_{\rm gw}$. The static pulsar parameters $\boldsymbol{\theta'}_{\rm psr}$ are specified in Section \ref{sec:rep_example1} and Table \ref{tab:parameters_and_priors}. It is prohibitive to display corner plots analogous to Figure \ref{fig:corner_plot_1} for 200 injections and seven elements of $\boldsymbol{\theta}_{\rm gw}$. Therefore, we summarise the results with the aid of a parameter-parameter (PP) plot \citep{doi:10.1198/106186006X136976}. A PP plot displays the fraction of injections included within a given credible interval of the estimated posterior, as a function of the credible interval itself. In the ideal case of perfect recovery, the PP plot should be a diagonal line of unit slope. \newline 
 
The PP results are displayed in Figure \ref{fig:parameter_space}. The shaded grey contours enclose the $1\sigma$, $2\sigma$, and $3\sigma$ significance levels for 200 injections. The curves for the five parameters, colour-coded in the legend, are approximately linear. Some parameters are consistently well estimated across the parameter domain, e.g. the blue curve for $\Omega$ lies everywhere within the 2$\sigma$ shaded region. Conversely some parameters, e.g. $\Phi_0$, stray outside the $1\sigma$ shaded region and show evidence of being slightly over-constrained; there are more injections contained within lower-value credible intervals than would be expected statistically, and fewer injections contained within higher-value credible intervals. This occurs due to the low SNR of the injected GW signal and the varying sensitivity of the specific PTA configuration as a function of sky position. More quantitatively, for the best estimated parameter $\Omega$ the injection is contained within the 90\% credible interval in 91\% of cases. For the worst estimated parameter $\Phi_0$, the injection is contained within the 90\% credible interval in 81\% of cases. The next paragraph explains why 81\% is less than the theoretical ideal 90\%.  \newline 

Manually inspecting the individual corner plots for each of the 200 injections, we see that the nested sampler does not return a unimodal posterior in the rare event that the source is located unfavourably, with $\boldsymbol{n} \cdot \boldsymbol{q}^{(n)} \approx -1$, such that one cannot infer $\chi^{(n)}$ accurately (see Section \ref{sec:chi_estim}). All injected sources are ``observed'' with the same settings such as $T_{\rm obs}$, $T_{\rm cad}$ and $n_{\rm live}$. In Figure \ref{fig:parameter_space2} we display a second PP plot, arranged identically to Figure \ref{fig:parameter_space}, but assuming that the injected values of $\chi^{(n)}$ are known exactly for the sake of testing, i.e. setting a delta-function prior on $\chi^{(n)}$. The excursions out of the shaded region diminish compared to Figure \ref{fig:parameter_space}, confirming the importance of estimating $\chi^{(n)}$ accurately. For the best estimated parameter $\Omega$ the injection is contained within the 90\% credible interval in 91\% of cases, the same as in Figure \ref{fig:parameter_space}. For the worst estimated parameters such as $\Phi_0$, the injection is contained within the 90\% credible interval in 89\% of cases, an improvement over the results of Figure \ref{fig:parameter_space}. \newline

\section{Bias when neglecting the pulsar terms}\label{sec:earth_vs_psr}
In Section \ref{sec:rep_example1} we validate the state-space analysis scheme when the pulsar terms are included in the inference model, i.e. Equation \eqref{eq:g_func_trig_chi}. In this section, we compare what happens when we omit the pulsar terms intentionally by using Equation \eqref{eq:g_func_trig_earth}. We do this for the specific representative case of the individual quasi-monochromatic SMBHB source described in Table \ref{tab:parameters_and_priors}. We refer to the model where the pulsar terms are included alongside the Earth term as $\mathcal{M}_{\rm psr \& Earth}$. We refer to the model where the pulsar terms are omitted as 
$\mathcal{M}_{\rm Earth}$. Model $\mathcal{M}_{\rm psr \& Earth}$ is parameterised by $\boldsymbol{\theta}'$. Model $\mathcal{M}_{\rm Earth}$ is parameterised by $\boldsymbol{\theta}_{\rm Earth} = \boldsymbol{\theta}_{\rm gw} \cup \boldsymbol{\theta}''_{\rm psr}$ where $\boldsymbol{\theta}''_{\rm psr}$ equals $\boldsymbol{\theta}'_{\rm psr}$ with $\chi^{(n)}$ removed. Both models are applied to identical realisations of the data, generated using the procedure in Section  \ref{sec:rep_example1}. The priors on the static parameters in each model are also identical, with the addition of a uniform prior on $\chi^{(n)}$ for $\mathcal{M}_{\rm psr \& Earth}$, as described in Appendix \ref{sec:set_priors}. \newline

In Section \ref{sec:psr_v_earth_pe} we compare the accuracy of the estimates of $\boldsymbol{\theta'}$ and $\boldsymbol{\theta}_{\rm Earth}$ returned by the respective models. In Section \ref{sec:psr_v_earth_bayes} we compare the minimum detectable GW strain for the two models by calculating the model evidence and comparing it to the evidence for a null model that does not contain a GW.

\subsection{Accuracy of parameter estimation}\label{sec:psr_v_earth_pe}

In this section we apply the Kalman filter in conjunction with nested sampling in order to infer the joint posterior distribution for the static parameters with and without the pulsar terms. For $\mathcal{M}_{\rm Earth}$ we apply the Kalman filter using Equation \eqref{eq:g_func_trig_earth} to return $\mathcal{L}(\boldsymbol{Y}| \boldsymbol{\theta}_{\rm Earth})$ and the nested sampler to estimate $p(\boldsymbol{\theta}_{\rm Earth} | \boldsymbol{Y})$. For $\mathcal{M}_{\rm psr \& Earth}$ we apply the Kalman filter using Equation \eqref{eq:g_func_trig_chi} to return $\mathcal{L}(\boldsymbol{Y}| \boldsymbol{\theta}')$ and the nested sampler to estimate $p(\boldsymbol{\theta'} | \boldsymbol{Y})$. The settings of the nested sampler (for example the number of live points; see Appendix \ref{sec:nested_sampling}) are identical for both models. \newline 

We consider two representative systems. The first is a ``low-SNR'' system with $h_0 = 5 \times 10^{-15}$, i.e. the system described in Table \ref{tab:parameters_and_priors}. The second is a ``high-SNR'' system which has the same static parameters as in Table \ref{tab:parameters_and_priors} except with $h_0 = 1 \times 10^{-12}$. The ``high-SNR'' system is considered in order to quantify any systematic biases, as distinct from random errors caused by the measurement noise. In order to enable a clear comparison between the posterior probability distributions calculated using the two models, in this section we present a single noise realisation of the synthetic data $\boldsymbol{Y}$. The results quoted are consistent across different noise realisations. \newline

\begin{figure*}
	\includegraphics[width=\textwidth, height =\textwidth ]{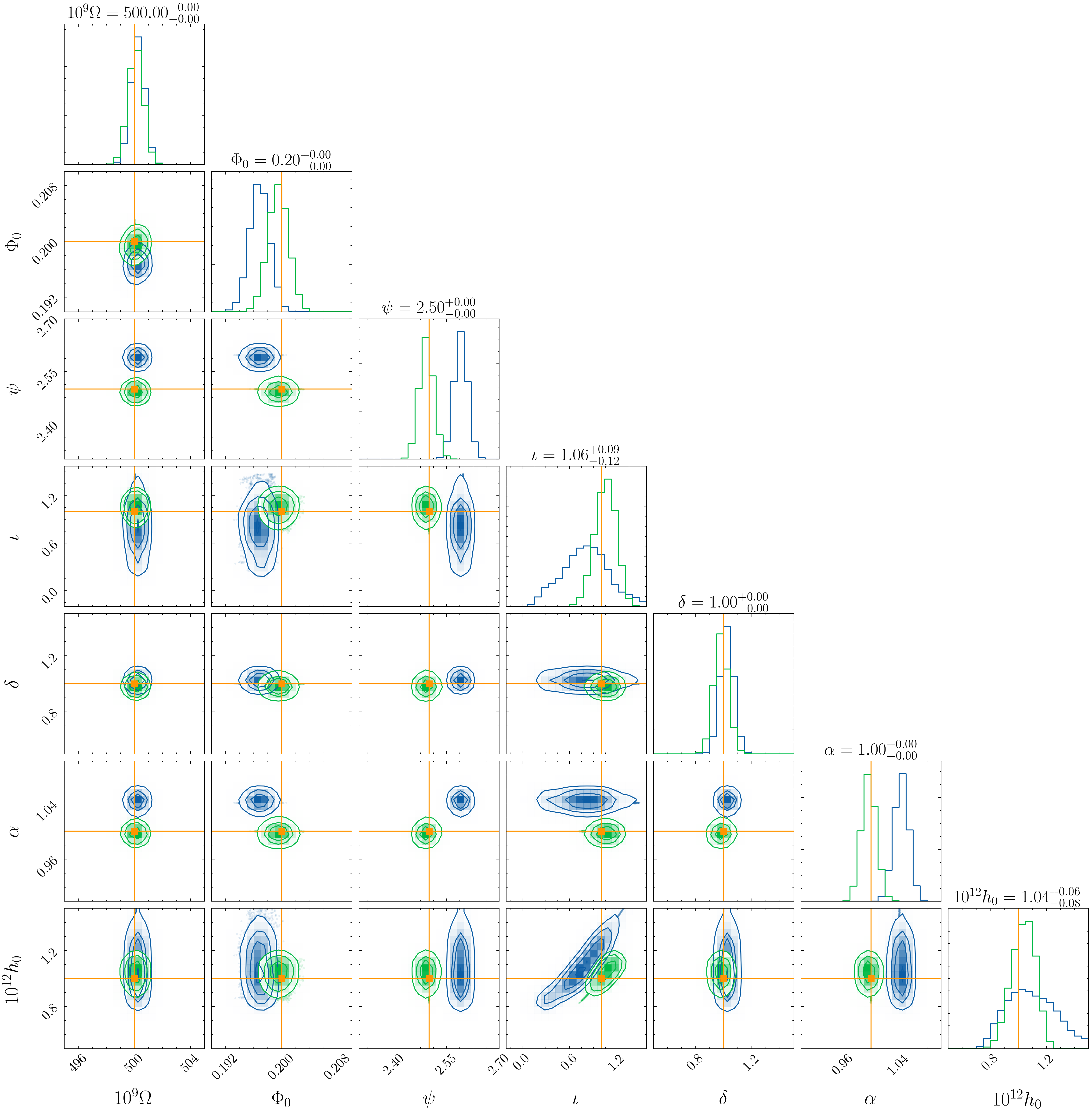}
	\caption{Posterior distribution in the form of a standard corner plot of the GW source parameters $\boldsymbol{\theta}_{\rm gw}$ for the representative system described in Table \ref{tab:parameters_and_priors}, with $h_0 = 1 \times 10^{-12}$, for a single realisation of the system noise. The blue curves are the posteriors calculated using the Earth-term model, Equation \eqref{eq:g_func_trig_earth}. The green curves are calculated by including the Earth term and pulsar terms, Equation \eqref{eq:g_func_trig}. The vertical and horizontal orange lines indicate the true injected values. The contours in the two-dimensional histograms denote the (0.5, 1, 1.5, 2)-$\sigma$ levels. The supertitles of the one-dimensional histograms record the medians and the 0.16 and 0.84 quantiles of the green curves, i.e. the pulsar-term model. We plot the scaled variables $10^9 \Omega$ (units: rad s$^{-1}$) and $10^{12} h_0$. Some parameters (e.g. $\psi, \iota$) exhibit a bias when using the Earth-term model, which disappears when the pulsar terms are included.}
	\label{fig:corner_plot_compare_high}
\end{figure*}

\begin{figure*}
	\includegraphics[width=\textwidth, height =\textwidth ]{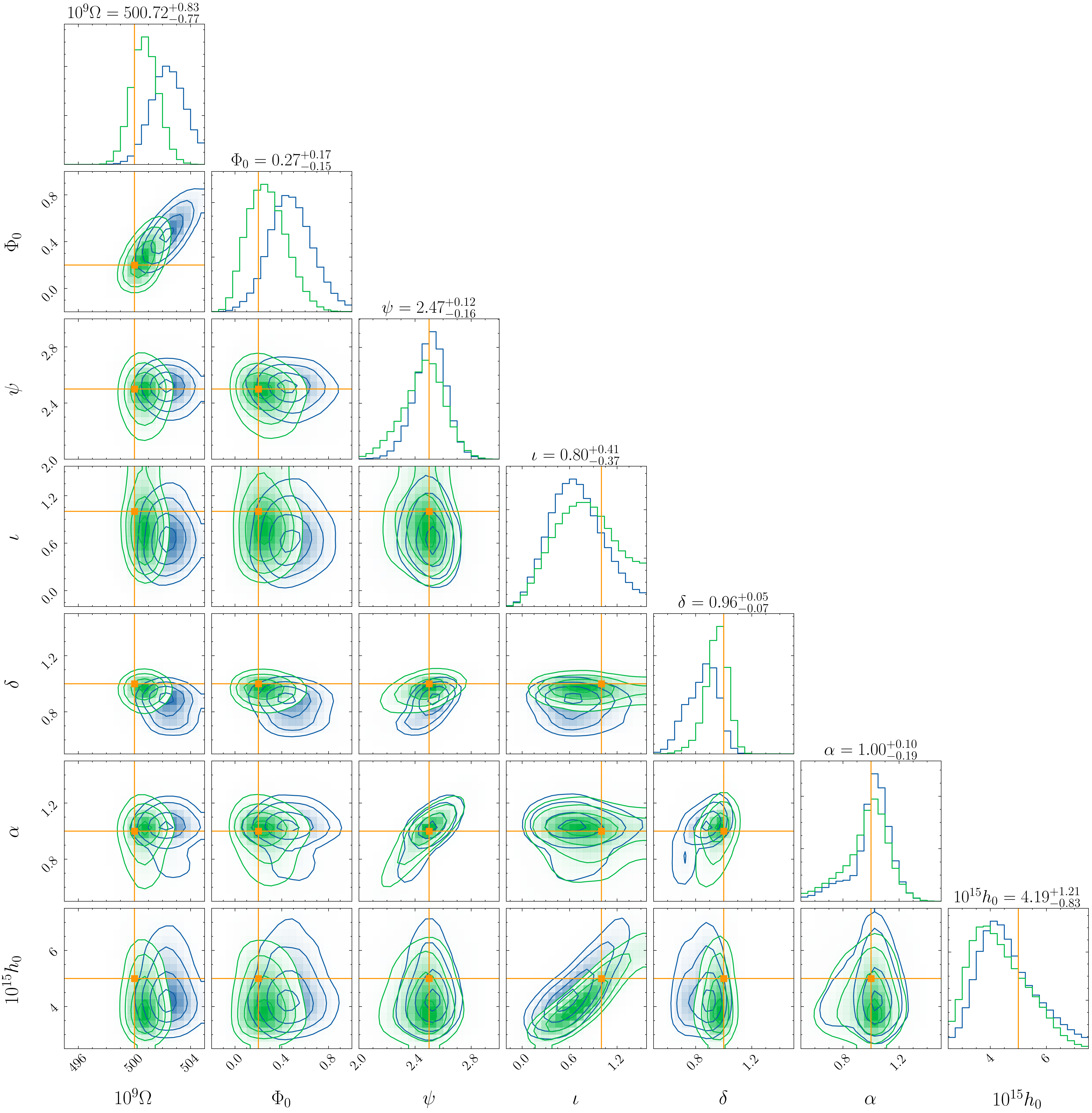}
	\caption{Same as Figure \ref{fig:corner_plot_compare_high}, but for a low-SNR system with $h_0 = 5 \times 10^{-15}$. The posterior distributions with (green curves) and without (blue curves) the pulsar terms overlap more closely than in Figure \ref{fig:corner_plot_compare_high}, but the green curves are still centered slightly more accurately on the injected values (horizontal and vertical orange lines).}
	\label{fig:corner_plot_compare_low}
\end{figure*}

The results for the seven parameters in $\boldsymbol{\theta}_{\rm gw}$ are shown in Figure \ref{fig:corner_plot_compare_high} for the high-SNR system and in Figure \ref{fig:corner_plot_compare_low} for the low-SNR system. The corner plot is arranged identically to Figure \ref{fig:corner_plot_1}, except that the different coloured curves now correspond to different inference models, rather than different realisations of $\boldsymbol{Y}$. The blue curves are the results derived using $\mathcal{M}_{\rm Earth}$. The green curves are the results derived using $\mathcal{M}_{\rm psr \& Earth}$. \newline 
		
For the high-SNR results in Figure \ref{fig:corner_plot_compare_high}, the one-dimensional posteriors for $\mathcal{M}_{\rm Earth}$ are biased, as was observed by K24, as well as \cite{Zhupulsarterms} and \cite{Chen2022}. Particular biases are observed in $\psi, \iota $ and $\alpha$, with the blue posterior displaced with respect to the orange injection line. For example, for $\iota$, the injected value is contained within the 90\% credible interval, but the median of the posterior of the Earth-term model is shifted from the injected value by 0.35 radians. The inclusion of the pulsar terms corrects for this bias. The green one-dimensional posteriors exhibit no bias and are generally symmetric about the injected value. Moreover, for every pair of parameters in ${\boldsymbol{\theta}}_{\rm gw}$, the injected values are contained within the 2-sigma contours for $\mathcal{M}_{\rm psr \& Earth}$. This is not true for $\mathcal{M}_{\rm Earth}$, where the injected values fall outside the 2-sigma contour for 15 out of the 21 parameter pairs.\newline

We define a relative error to quantify the accuracy of the one-dimensional posteriors with respect to the injected value. The relative error is defined to be the relative unsigned displacement of the mode of $p_{\rm M}\left(\theta| \boldsymbol{Y}\right)$ from the injection, viz.
\begin{eqnarray}
\Delta_{\rm M}(\theta) = \frac{ \left| \underset{\theta}{\text{argmax }} p_{\rm M}\left(\theta| \boldsymbol{Y}\right) - \theta_{\rm inj}\right|}{\theta_{\rm inj}} \ . \label{eq:mean_rel_error}
\end{eqnarray}
In Equation \eqref{eq:mean_rel_error}, $p_{\rm M}\left(\theta| \boldsymbol{Y}\right)$ is the one-dimensional posterior for parameter $\theta$ returned by nested sampling (c.f. Equation \eqref{eq:posterior_distrib}). The subscript $\rm M \in \{ \text{Earth, psr\&Earth} \}$ indicates whether the posterior is estimated using Equation \eqref{eq:g_func_trig_earth} or Equation \eqref{eq:g_func_trig} respectively. $\theta_{\rm inj}$ denotes the true injection value. We note that $\Delta_{\rm M}(\theta)$ quantifies the accuracy of the estimates returned by the two models, i.e. the closeness of the most probable estimate of $\theta$ relative to $\theta_{\rm inj}$. In contrast, $\Delta_{\rm M}(\theta)$ does not quantify the uncertainty in the estimates. The error $\Delta_{\rm M}(\theta)$ for each parameter in Figure \ref{fig:corner_plot_compare_high} is summarised in Table \ref{tab:posterior_errors}. We find that the estimates are more accurate using the pulsar terms, i.e. $\Delta_{\rm Earth}(\theta) > \Delta_{\rm psr \& Earth}(\theta)$ for $\theta \in \boldsymbol{\theta}_{\rm gw}$. In some cases the difference is modest;  $\Omega$ is recovered with high accuracy by both models, with $\Delta_{\rm Earth}(\Omega)-\Delta_{\rm psr \& Earth}(\Omega) = 1.5 \times 10^{-5}$. The improvement from including the pulsar terms is largest for $\iota$, with $\Delta_{\rm Earth}(\iota) - \Delta_{\rm psr \& Earth}(\iota)  = 0.14$. Whilst we present only a single noise realisation in this section, the improvements from including the pulsar terms are found to be comparable across different realisations. \newline

For the low-SNR results in Figure \ref{fig:corner_plot_compare_low}, there is less improvement in the parameter estimates. Qualitatively, the green and blue contours and histograms mostly overlap, although the green contours are centred slightly better on the injected values. The relative error $\Delta_{\rm M}(\theta)$ for the low-SNR results is reported in the lower half of Table \ref{tab:posterior_errors}. We see that the error is larger for every element in $\boldsymbol{\theta}_{\rm gw}$ than in the high-SNR case, as expected. The inclusion of the pulsar terms improve the estimates for five out of the seven static parameters; we find $\Delta_{\rm Earth}(\theta) > \Delta_{\rm psr \& Earth}(\theta)$ for all parameters except $\psi$ and $h_0$. However it is hard to draw strong conclusions; the improvements from including the pulsar terms vary randomly across different noise realisations. We show in Section \ref{sec:psr_v_earth_bayes} that including the pulsar terms increases the detection probability. \newline 

\begin{table}
	\centering
		\begin{tabular}{lcll}
			\toprule
			&$\theta$ & $\Delta_{\rm Earth}(\theta)$ & $\Delta_{\rm psr \& Earth}(\theta)$  \\
			\hline
			\multirow{7}{2mm}{High SNR} & $\Omega$       & $1.9 \times 10^{-5}$& $3.4 \times 10^{-6}$\\
			& $\Phi_0$ &$1.6 \times 10^{-2}$ &$2.2 \times 10^{-3}$ \\
			& $\psi$ &$3.9 \times 10^{-2}$ &$2.0 \times 10^{-4}$  \\
			& $\iota$ & $2.0 \times 10^{-1}$ & $6.1 \times 10^{-2}$  \\
			& $\delta$ &$1.9 \times10^{-3}$ &$2.0 \times 10^{-4}$  \\
			&$\alpha$ &$4.0 \times 10^{-2}$ &$2.8 \times10^{-4}$  \\
			&$h_0$ & $9.3 \times 10^{-2}$ &$4.9 \times 10^{-2}$  \\
			\hline
			\multirow{7}{2mm}{Low SNR} & $\Omega$       &  $4.8 \times 10^{-3}$ &$1.5 \times 10^{-3}$ \\
			& $\Phi_0$ &$1.4 \times 10^{0}$ & $3.3 \times 10^{-1}$  \\
			& $\psi$ &$5.3 \times 10^{-3}$ &$1.1 \times 10^{-2}$  \\
			& $\iota$ & $3.3 \times 10^{-1}$ & $2.1 \times 10^{-1}$  \\
			& $\delta$ & $1.3 \times 10^{-1}$ & $4.2 \times 10^{-2}$  \\
			&$\alpha$ & $2.5 \times 10^{-2}$ & $3.3 \times 10^{-3}$ \\
			&$h_0$ & $1.1 \times 10^{-1}$ &$1.6 \times 10^{-1}$ \\
			\bottomrule
		\end{tabular}
		\caption{Relative error $\Delta_{\rm M}(\theta)$, Equation \eqref{eq:mean_rel_error}, in the mode of the one-dimensional posteriors calculated using the Earth-term model (M = Earth, Equation \eqref{eq:g_func_trig_earth}) and the pulsar-term model (M = $\rm psr \& Earth$, Equation \eqref{eq:g_func_trig}) for $\theta \in \boldsymbol{\theta}_{\rm gw}$. The injected values are summarised in Table \ref{tab:parameters_and_priors}. The top and bottom halves of the table contain the high-SNR ($h_0 =1 \times 10^{-12}$) and low-SNR ($h_0 = 5 \times 10^{-15}$) cases respectively. The psr\&Earth model is more accurate than the Earth model for all $\theta \in \boldsymbol{\theta}_{\rm gw}$ at high SNR, and five out of seven $\theta \in \boldsymbol{\theta}_{\rm gw}$ at low SNR.}
		\label{tab:posterior_errors}
	\end{table}

\subsection{Detectability vs $h_0$}\label{sec:psr_v_earth_bayes}
		\begin{figure}
			\includegraphics[width=\columnwidth, height = \columnwidth ]{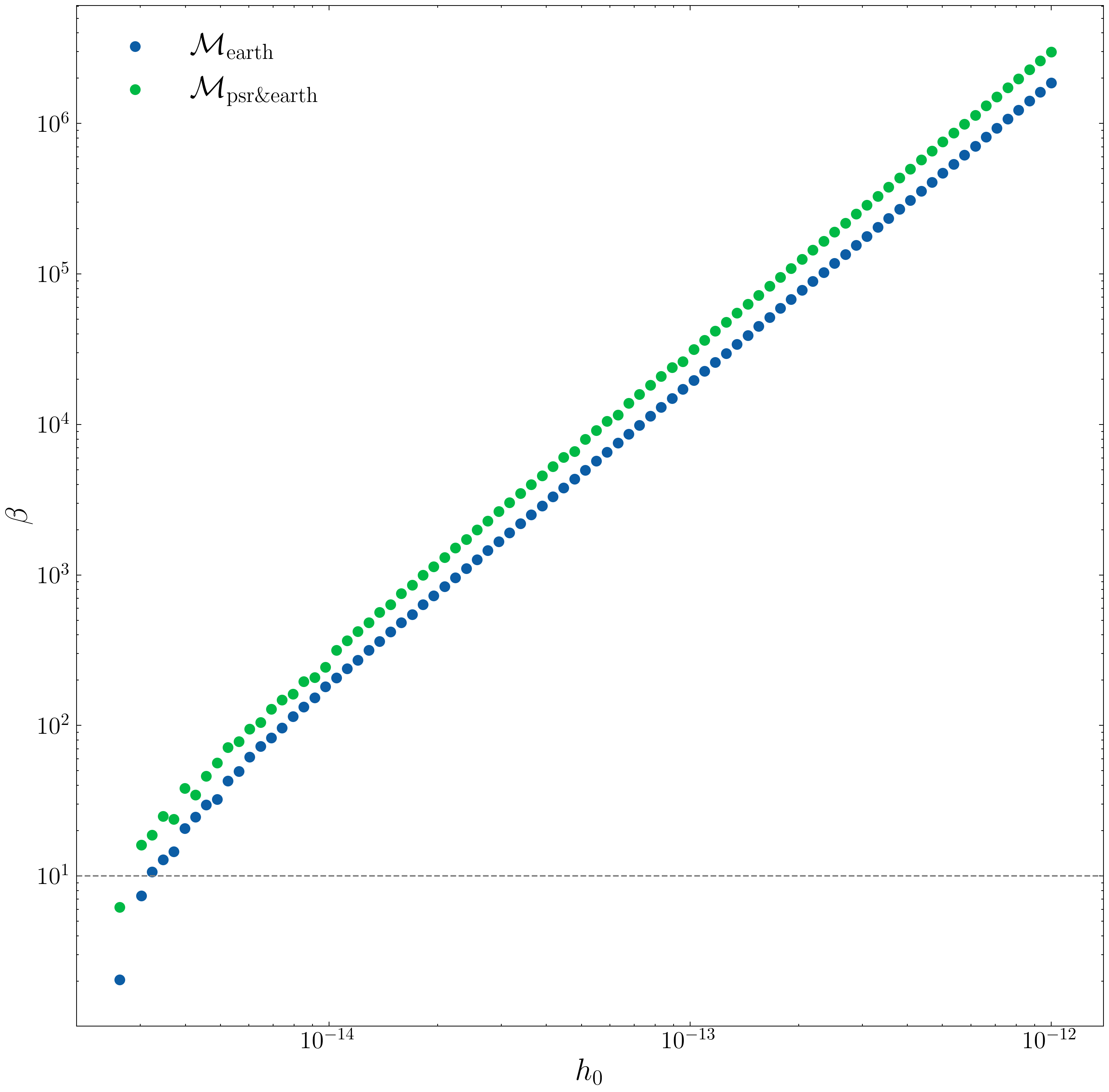} 	
			\caption{Bayes factor (odds ratio) $\beta_{\rm M}$ between the competing models $\mathcal{M}_{\rm M}$, with $\rm M \in \left \{ \rm psr \& Earth, Earth \right \}$ (GW present in data) and $\mathcal{M}_{\rm null}$ (GW not present in data) as a function of the signal amplitude, $h_0$, for the representative example in Table \ref{tab:parameters_and_priors}. The horizontal grey dashed line labels an arbitrary detection threshold, $\beta_{\rm M} = 10$. The minimum detectable strain at $\beta_{\rm M} = 10$ equals $3.2 \times 10^{-15}$ for $\mathcal{M}_{\rm Earth}$ (blue points) and $2.8 \times 10^{-15}$ for $\mathcal{M}_{\rm psr \& Earth}$ (green points). The axes are plotted on logarithmic scales.}
			\label{fig:bayes1}
		\end{figure}
	
In this section we compute the minimum detectable strain for the representative source in Table \ref{tab:parameters_and_priors}, using $\mathcal{M}_{\rm Earth}$ and $\mathcal{M}_{\rm psr \& Earth}$. We frame the detection problem in terms of a Bayesian model selection procedure, following the lead of other PTA analyses \citep[e.g.][]{2023ApJ...951L...8A,2023arXiv230616214A,2023ApJ...951L...6R,2023RAA....23g5024X}. We define $\mathcal{M}_{\rm null}$ as the null model that assumes no GW exists in the data. This is equivalent to setting $g^{(n)}(t)=1$ in Equation \eqref{eq:g_func_trig}. The evidence integral $\mathcal{Z}$ returned by nested sampling, Equation \eqref{eq:model_evidence}, is the probability of the data $\boldsymbol{Y}$ given a model $\mathcal{M}_{\rm M}$. The support in the data for the presence of a GW signal, described by model $\mathcal{M}_{\rm M}$, over the absence of a GW signal is quantified via the Bayes factor
		\begin{equation}
			\beta_{\rm M} = \frac{\mathcal{Z}(\boldsymbol{Y} | \mathcal{M}_{\rm M})}{\mathcal{Z}(\boldsymbol{Y} | \mathcal{M}_{\rm null})} \ . \label{eq:bayes}
		\end{equation}
In this paper we consider $\rm M \in \left \{ \rm Earth, psr \& Earth \right \}$. \newline 
	
The Bayes factors $\beta_{\rm Earth}$ and $\beta_{\rm psr{\&}Earth}$ are plotted as functions of $h_0$ in Figure \ref{fig:bayes1} for the representative source in Table \ref{tab:parameters_and_priors}. We vary the source amplitude from $h_0 = 10^{-15}$ (undetectable) to $h_0 = 10^{-12}$ (easily detectable). To sharpen the comparison between $\mathcal{M}_{\rm psr \& Earth}$ and $\mathcal{M}_{\rm Earth}$ we present only a single noise realisation of the synthetic data $\boldsymbol{Y}$, as in Section \ref{sec:psr_v_earth_pe}; the conclusions drawn below are consistent across different noise realisations. Moreover, the noise processes in the synthetic data are identical realisations for each value of $h_0$; the only change from one $h_0$ value to the next is $h_0$ itself, to smooth the curves in Figure \ref{sec:psr_v_earth_pe}. \newline

Figure \ref{fig:bayes1} reveals an approximate quadratic relationship $\beta \propto h_0^2$ for $h_0 \gtrsim 10^{-14}$ for both $\mathcal{M}_{\rm Earth}$ and $\mathcal{M}_{\rm psr \& Earth}$. Moreover, we obtain $\beta_{\rm psr \& Earth} > \beta_{\rm Earth}$ for all $h_0$. That is, for a given $h_0$, $\mathcal{M}_{\rm psr \& Earth}$ provides greater evidence for a GW signal in the noisy data than $\mathcal{M}_{\rm Earth}$. The GW source is detectable with decisive evidence ($\beta \geq 10$) for $h_0 \gtrsim 3.2 \times 10^{-15}$ for  $\mathcal{M}_{\rm Earth}$ and $h_0 \gtrsim 2.8 \times 10^{-15}$ for  $\mathcal{M}_{\rm psr \& Earth}$, a relative improvement in the minimal detectable strain of $14\%$. The minimum detectable strain and the relative improvement through using $\mathcal{M}_{\rm psr \& Earth}$ are particular to the system in Table \ref{tab:parameters_and_priors} and the realisation of $\boldsymbol{Y}$. They are influenced in general by $T_{\rm obs}$, $T_{\rm cad}$, ${\boldsymbol{\theta}}_{\rm gw}$, and ${\boldsymbol{\theta}}_{\rm psr}$. A full parameter sweep is postponed to future work, after the analysis scheme in this paper is upgraded from ingesting pulse frequencies to pulse TOAs to enable a like-for-like comparison with standard PTA analyses. In the interim, we note that a $14\%$ improvement in sensitivity when including the pulsar terms is comparable to the improvements of $5\%$ achieved by \cite{Zhupulsarterms}.

\section{Conclusion}\label{sec:discussion}
In this paper we demonstrate how to extend state-space methods for PTA data analysis to include the pulsar terms as well as the Earth term. We emphasize that including the pulsar terms is not a new idea; the advantages of doing so are well known in standard PTA analyses 
\citep[e.g.][]{Zhupulsarterms,Chen2022,2023ApJ...951L...8A,2023arXiv230616214A,2023ApJ...951L...6R,2023RAA....23g5024X,Arzoumanian2023,2023arXiv230616226A}. The goal of this paper is to verify whether the advantages apply equally to state-space methods, which complement standard analyses. In the state-space formulation, the rotational state of each pulsar evolves according to a mean-reverting Ornstein-Uhlenbeck process, Equations \eqref{eq:frequency_evolution}--\eqref{eq:spinevol}, and is tracked using a Kalman filter. The measurement equation in the Kalman filter is reparameterized in terms of a pulsar-specific static phase $\chi^{(n)}$ ($1 \leq n \leq N$) to be inferred for each pulsar. The Kalman filter is combined with a nested sampler to estimate the posterior distributions of each static parameter, as well as the associated Bayesian evidence of the signal models with and without the pulsar terms included. \newline

The updated state-space model including the pulsar terms is tested on synthetic data. We start by considering 10 noise realisations for a single, astrophysically representative, SMBHB GW source with $h_0 = 5 \times 10^{-15}$, observed by the 12.5-year NANOGrav pulsars ($N=47$) with $T_{\rm obs} = 10 \, {\rm years}$ and $T_{\rm cad} = 1$ week. We find that the updated state-space model successfully detects injected signals and estimates their static parameters accurately for all noise realisations with relatively low computational cost. The irreducible random dispersion (cosmic variance) in the median of the marginalised one-dimensional posteriors is non-negligible, with a maximum $CV$ of 52\% for $\Phi_0$, a minimum of 0.2\% for $\Omega$ and a median of $10 \%$ for $h_0$. \newline 

The updated model is further tested across a broad and astrophysically plausible parameter domain of SMBHB sources, exploring 200 randomly sampled $\boldsymbol{\theta}_{\rm gw}$ at fixed $\iota$ and $h_0$. Consistent parameter estimates are obtained, although the accuracy is lower, when one or more PTA pulsars coincide approximately on the sky with the SMBHB source. For the best estimated parameter $\Omega$ the injection is contained within the 90\% credible interval in 91\% of cases. For the worst estimated parameter $\Phi_0$, the injection is contained within the 90\% credible interval in 81\% of cases. Including the pulsar terms increases the accuracy (as quantified by the root-mean-square relative error, Equation \eqref{eq:mean_rel_error}) for five out of the seven static parameters in $\boldsymbol{\theta}_{\rm gw}$; we find $\Delta_{\rm Earth}(\theta) > \Delta_{\rm psr \& Earth}(\theta)$ for all parameters except $\psi$ and $h_0$. In the high-SNR case ($h_0 = 1 \times 10^{-12}$) all of the seven static parameters are estimated more accurately; we find $\Delta_{\rm Earth}(\theta) > \Delta_{\rm psr \& Earth}(\theta)$ for $\theta$ in $\boldsymbol{\theta}_{\rm gw}$. Including the pulsar terms lowers the minimum detectable strain by $14\%$, comparable with standard PTA analyses \citep[e.g.][]{Zhupulsarterms}. \newline

State-space methods for PTA data analysis are complementary to traditional approaches. They track the actual, astrophysical, time-ordered realization of the timing noise in every PTA pulsar instead of fitting for the noise power spectral density (PSD), which averages over the ensemble of possible timing noise realizations (not just the actual one) and discards time-ordered information when taking the modulus of the complex Fourier phase in each PSD frequency bin. That is, it is possible to disentangle statistically the specific time-ordered realisation of the timing noise from the GW-induced modulations, and thereby infer the GW source parameters. State-space models are conditional on a noise model that is related but different to the noise model in traditional approaches. The analysis in this paper assumes a mean-reverting Ornstein-Uhlenbeck process, whereas traditional analyses assume a stationary Gaussian process described by an ensemble-averaged, power-law PSD, whose amplitude and exponent are adjustable. The Ornstein-Uhlenbeck process also maps onto a stationary Gaussian process, whose PSD is a power law at high frequencies and rolls over at low frequencies, and whose amplitude and exponent can be adjusted by modifying the form of the damping term and Langevin driver in Equation \eqref{eq:frequency_evolution}. However, the Ornstein-Uhlenbeck process in the time domain in Equation \eqref{eq:frequency_evolution} contains more information than its associated PSD in the frequency domain for the two reasons stated above: the Kalman filter ``fits'' the actual noise realization rather than an ensemble average, and it preserves time ordering by implicitly preserving the Fourier phases, which the PSD discards. Clarifying the similarities and differences between various approaches promises to be a fruitful avenue of future work. It is also a subject of attention in audio-band GW data analysis involving hidden Markov models applied to data from terrestrial long-baseline interferometers \citep{PhysRevD.102.023006,PhysRevD.105.022002,Abbott_2022SCO,2022PhRvD.106f2002A}. \newline 

There are at least four useful extensions to this work.

\begin{enumerate}[leftmargin=2em]
		
	\item For the single representative SMBHB source in Section \ref{sec:psr_v_earth_pe} there is no appreciable improvement in the parameter estimation accuracy from including the pulsar terms for $h_0 = 5\times 10^{-15}$ (i.e.\ low SNR). It would be of interest to explore a broader $\boldsymbol{\theta}'$ domain (i.e. varying both the SMBHB source and the PTA configuration) to check whether including the pulsar terms improves the estimation accuracy at low SNR in some overlooked pockets of the astrophysically plausible ${\boldsymbol{\theta}}'$ domain. A similar exercise should be undertaken with respect to the minimum detectable GW strain (see Section \ref{sec:psr_v_earth_bayes}  ). \newline 
	
	\item The assumption of a monochromatic source is well-justified astrophysically in various regimes (see Section \ref{sec:psr_measured}).  Nevertheless, SMBHBs are not strictly monochromatic. In principle it is straightforward to include the evolution of $\Omega(t)$ in the differential state equations. However, issues related to identifiability \citep{e5be7c83a0d24500826f6e1b414d1733}, the form of Equation \eqref{eq:g_func_trig}, and the linear structure of the Kalman filter must be considered carefully and are postponed to a future paper. The Kalman framework can be applied to non-linear problems if needed using either an extended Kalman filter \citep{zarchan2000fundamentals}, unscented Kalman filter \citep{882463van} or particle filter \citep{Simon10}. Preliminary performance tests regarding the monochromatic assumption are presented in Appendix \ref{sec:monochromatic}. \newline

	\item The state-space analysis presented here ingests a frequency time series $f_{\rm m}^{(n)}(t)$. It is necessary to generalize this approach to ingest TOAs directly, as happens in standard PTA analyses \citep[e.g.][]{Zhupulsarterms,Chen2022,2023ApJ...951L...8A,2023arXiv230616214A,2023ApJ...951L...6R,2023RAA....23g5024X,Arzoumanian2023,2023arXiv230616226A}. Generalizing the algorithm is a surprisingly subtle task and will be presented in a forthcoming paper. \newline

	\item We assume in this paper that there is only one GW source. The Kalman framework extends naturally to multiple sources. It is straightforward to modify Equation \eqref{eq:measurement} to describe a linear superposition of GWs. This is useful for two reasons. Firstly, it may be possible to resolve multiple continuous GW sources concurrently \citep{PhysRevD.85.044034}. Secondly, the stochastic background itself is an incoherent sum of many individual GW sources. It should be possible for a Kalman filter and nested sampler to operate together to detect the stochastic background. With respect to the first reason, it is straightforward to apply the Kalman filter and nested sampler with an extended range of static parameters associated with the GW sources. With respect to the second reason, it is necessary to summarize economically the additional static parameters, whilst respecting the mathematical structure of the Kalman filter. This is a subtle challenge which we postpone to a forthcoming paper. If successful, it will complement the traditional approach of cross-correlating TOA residuals to uncover the Hellings-Downs curve \citep{Hellings,2023ApJ...951L...8A}.
\end{enumerate}

\section*{Acknowledgements}
This research was supported by the Australian Research Council Centre of Excellence for Gravitational Wave Discovery (OzGrav), grant number CE170100004. The numerical calculations were performed on the OzSTAR supercomputer facility at Swinburne University of Technology. The OzSTAR program receives funding in part from the Astronomy National Collaborative Research Infrastructure Strategy (NCRIS) allocation provided by the Australian Government.

\section*{Data Availability}
No new data were generated or analysed in support of this research.

\bibliographystyle{mnras}
\bibliography{example} 

\appendix
\newpage
\newpage
\clearpage

\section{Validity of the monochromatic assumption}\label{sec:monochromatic}
In this paper, as discussed in Section \ref{sec:psr_measured}, we treat the GW source as non-evolving, such that $\Omega$ is constant, and the GW that modulates the received pulsar signal is monochromatic. This is a reasonable approximation for the primary goal of this paper, namely studying the biases incurred by omitting the pulsar terms within a state-space formulation. Previous investigations of pulsar-term biases in the context of standard PTA analyses \citep[e.g.][]{Zhupulsarterms,Chen2022} also treat the GW source as non-evolving. \newline

The non-evolving approximation is almost exact for SMBHBs over the timescale set by $T_{\rm obs} \sim 10$ years; see Equation \eqref{eq:f_evolution} and the associated discussion in Section \ref{sec:psr_measured}. However, over timescales that correspond to the light travel time between pulsar and Earth (i.e. $d^{(n)}/c \gg T_{\rm obs}$) the inspiralling binary typically evolves appreciably. Consequently the value of $\Omega$ for the GW as it strikes the Earth (i.e. the Earth term) is distinct from the value of $\Omega$ as it strikes a pulsar (i.e. the pulsar term). In this appendix we repeat the Bayesian inference analysis of the main text (Section \ref{sec:pe_and_ms}) but now consider synthetic data for evolving sources, whilst retaining the non-evolving inference model (Section \ref{sec:psr_measured}). Our purpose is not to perform an in-depth performance analysis, but rather to confirm that the conclusions drawn in the main text are not compromised substantially by the treating the GW source as non-evolving.  This appendix is organised as follows. In Appendix \ref{app:validation_chromatic} we review how the value of $\Omega$ at the pulsar is related to the value of $\Omega$ measured at Earth. We outline how to generate synthetic data for evolving sources and the procedure for testing the performance of the method on the new synthetic data. In Appendix \ref{sec:chromatic_pe} we apply the state-space analysis scheme of the main text to synthetic data for evolving sources. We recover the system parameters $\boldsymbol{\theta}_{\rm gw}$, and compare the accuracy of the results with those obtained for non-evolving sources. In Appendix \ref{app:detection_chromatic} we compare the minimum detectable GW strain for evolving sources with that of non-evolving sources. \newline

\subsection{Quasi-monochromatic synthetic data}\label{app:validation_chromatic}

The angular frequency of the GW at the Earth, $\Omega_{\rm Earth}$, is related to the angular frequency of the GW at the $n$-th pulsar, $\Omega_{\rm psr}^{(n)}$, by \citep{Perrodin2018,2023ApJ...951L..50A,Arzoumanian2023}
\begin{equation}
	\Omega_{\rm psr}^{(n)} = \Omega_{\rm Earth} \left \{1 + \frac{256}{5} M_{\rm c}^{5/3} \Omega_{\rm earth}^{8/3} \left [t_{\rm p}^{(n)} - t \right] \right \}^{-3/8} \, ,\label{eq:omega_evol}
\end{equation}
where $t_{\rm p}^{(n)}$ is the retarded time at which the GW is incident on the $n$-th pulsar, $t$ is the time at which the GW is incident on the Earth, and we have
\begin{equation}
	t_{\rm p}^{(n)} - t = - d^{(n)} \left[1 +  \boldsymbol{n} \cdot \boldsymbol{q}^{(n)}\right] \, .
\end{equation}
Note that $\Omega_{\rm psr}^{(n)}$ depends on $n$ and is smaller than $\Omega_{\rm Earth}$. For example, for a GW with $\Omega_{\rm Earth} = 5 \times 10^{-7}$ Hz, Equation \eqref{eq:omega_evol} implies that a pulsar at a distance of 1 kpc experiences $\Omega_{\rm psr}^{(n)} \approx 3 \times 10^{-7}$ Hz, taking $M_{\rm c} = 10^8 M_{\odot}$. \newline

In Sections \ref{sec:2}--\ref{sec:earth_vs_psr} we assume $\Omega_{\rm psr}^{(n)} = \Omega_{\rm Earth}$ for all $n$. If we relax this assumption, the measurement equation, Equation \eqref{eq:g_func_trig_chi}, becomes \citep{Perrodin2018}
\begin{align}
	g^{(n)}_{\rm evolving}(t) =& 1 - \frac{ H_{ij}[q^{(n)}]^i [q^{(n)}]^j }{2 [1 + \boldsymbol{n}\cdot \boldsymbol{q}^{(n)}] } \nonumber \\
	& \times \Big \{\cos\left(-\Omega_{\rm earth} t +\Phi_0\right) \nonumber \\
	&- \cos \left [-\Omega^{(n)}_{\rm psr} t +\Phi_0 + \chi^{(n)} \right ] \Big \} \ .
	\label{eq:g_func_trig_chi_chromatic}
\end{align}
In order to test how sensitive the results in Sections \ref{sec:2}--\ref{sec:earth_vs_psr} are to the assumption $\Omega_{\rm psr}^{(n)} = \Omega_{\rm Earth}$, we take the following steps.
\begin{enumerate}[leftmargin=2em]
	\item Generate synthetic data $\boldsymbol{Y}_{\rm evolving}$ using Equation \eqref{eq:g_func_trig_chi_chromatic} via the procedure outlined in Section \ref{sec:rep_example1}, using the parameters defined in Table \ref{tab:parameters_and_priors}.
	\item Pass $\boldsymbol{Y}_{\rm evolving}$ into the Kalman filter - and nested sampler described in Section \ref{sec:pe_and_ms}. The analysis pipeline retains Equation \eqref{eq:g_func_trig_chi} as the measurement equation for the purposes of inference.
	\item Calculate $p({\boldsymbol{\theta}'} | {\boldsymbol{Y}_{\rm evolving}})$.
	\item Calculate the evidences $\mathcal{Z}(\boldsymbol{Y}_{\rm evolving} | \mathcal{M}_{\rm \rm psr \& Earth})$ and $\mathcal{Z}(\boldsymbol{Y}_{\rm evolving} | \mathcal{M}_{\rm null})$.
	\item Take the ratio of the evidences in step (iv) to obtain a Bayes factor, $\beta_{\rm psr \& Earth, evolving}$ cf. Equation \eqref{eq:bayes}.
\end{enumerate}
In step (i), we assume that the SMBHB has chirp mass $M_{\rm c} = 10^{8} M_{\odot}$. Step (iii) is the focus of Appendix \ref{sec:chromatic_pe}. We compare $p({\boldsymbol{\theta}'} | {\boldsymbol{Y}_{\rm evolving}})$ and $p({\boldsymbol{\theta}'} | {\boldsymbol{Y}})$, the posterior obtained in the main text for monochromatic data. Steps (iv) and (v) are the focus of Appendix \ref{app:detection_chromatic}. We compare $\beta_{\rm psr \& Earth, evolving}$ with $\beta_{\rm psr \& Earth}$, the Bayes factors obtained in the main text for monochromatic data.

\subsection{Parameter estimation}\label{sec:chromatic_pe}
\begin{figure*}
	\includegraphics[width=\textwidth, height =\textwidth ]{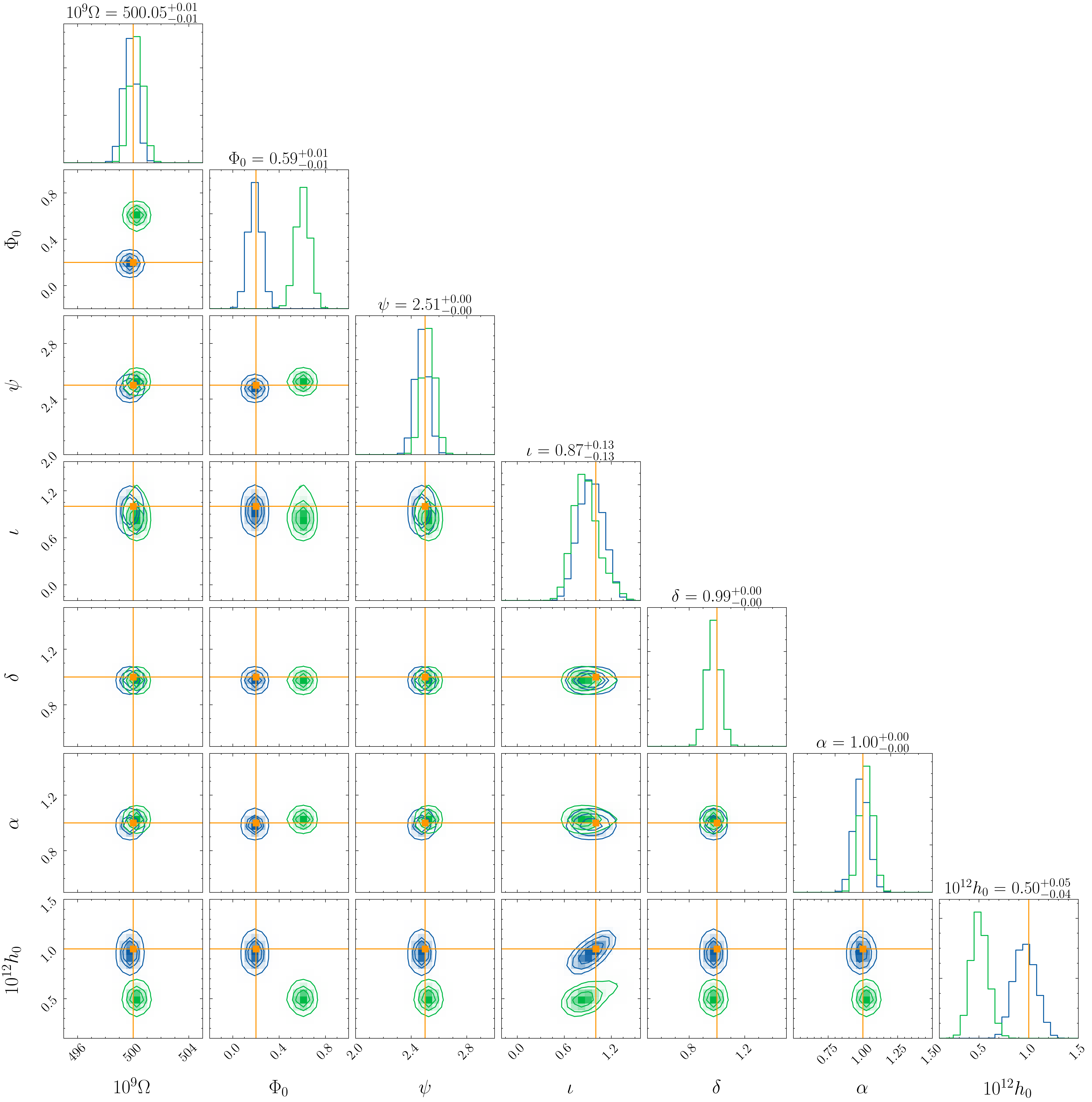}
	\caption{Impact on parameter estimation of inspiral-driven SMBH evolution on the GW frequency at Earth and at the retarded time at every pulsar. Posterior distribution in the form of a standard corner plot of the GW source parameters $\boldsymbol{\theta}_{\rm gw}$ for the representative system described in Table \ref{tab:parameters_and_priors}, with $h_0 = 1 \times 10^{-12}$, for a single realisation of the system noise. The blue curves are the posteriors calculated using the Earth and pulsar terms inference model, Equation \eqref{eq:g_func_trig}, computed for non-evolving data $\boldsymbol{Y}$. The green curves are the posteriors calculated using the same inference model for evolving data $\boldsymbol{Y}_{\rm evolving}$. The vertical and horizontal orange lines indicate the true injected values. The contours in the two-dimensional histograms denote the (0.5, 1, 1.5, 2)-$\sigma$ levels. The supertitles of the one-dimensional histograms record the medians and the 0.16 and 0.84 quantiles of the green curves. We plot the scaled variables $10^9 \Omega$ (units: rad s$^{-1}$) and $10^{12} h_0$. Qualitatively, the posteriors are broadly similar, with modest shifts in the modes of $\Phi_0$ and $h_0$.}
	\label{fig:monochromatic_assumption_high_snr}
\end{figure*}

\begin{figure*}
	\includegraphics[width=\textwidth, height =\textwidth ]{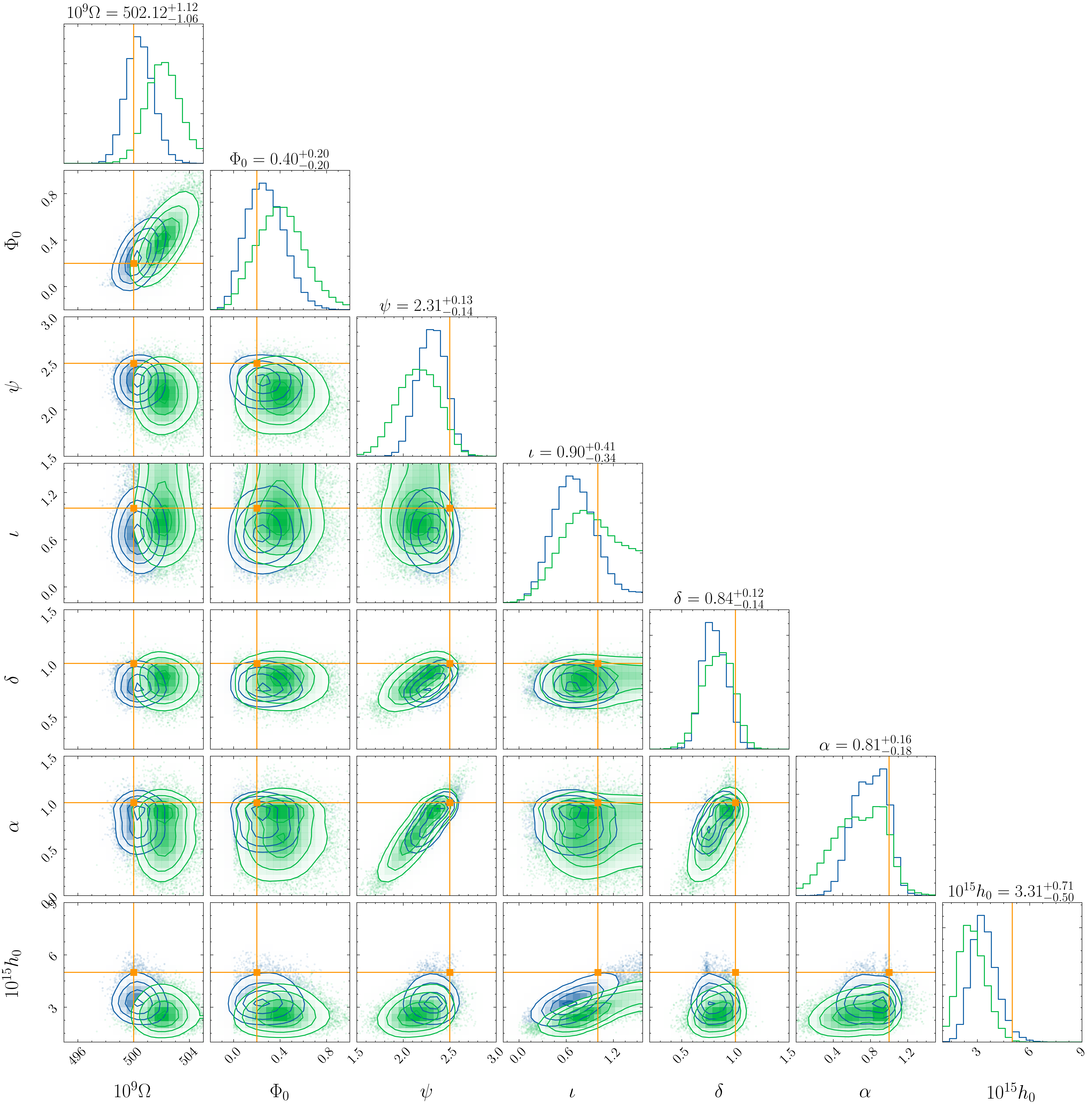}
	\caption{Same as Figure \ref{fig:monochromatic_assumption_high_snr}, but for a low-SNR system with $h_0 \ 5 \times 10^{-15}$. Qualitatively, the blue and green curves are similar and overlap for all $\theta \in \boldsymbol{\theta}_{\rm gw}$. }
	\label{fig:monochromatic_assumption_low_snr}
\end{figure*}

\begin{figure}
	\includegraphics[width=\columnwidth, height =\columnwidth ]{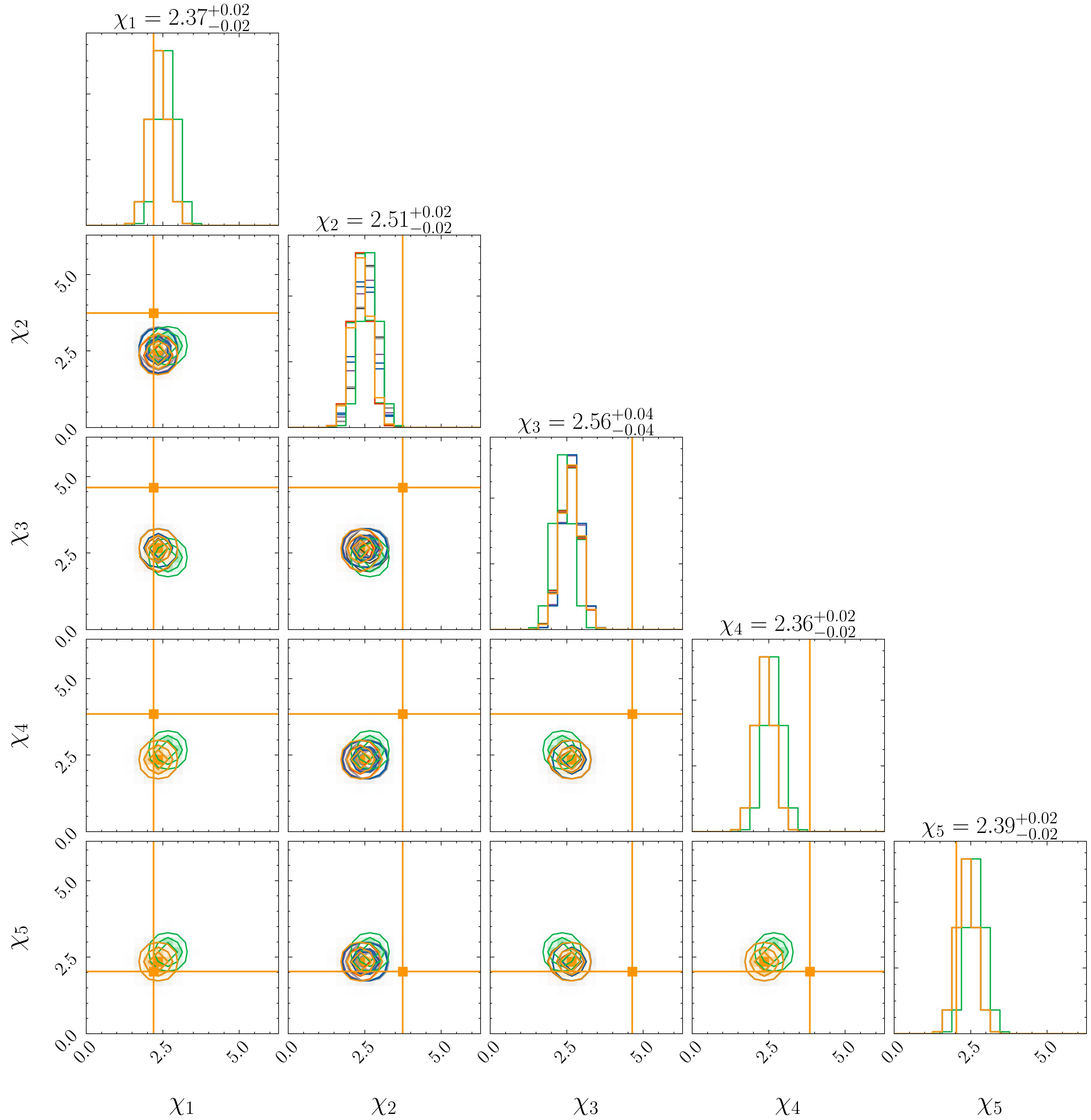}
	\caption{Same as Figure \ref{fig:corner_plot_3}, but for synthetic data $\boldsymbol{Y}_{\rm evolving}$ generated for an inspiralling SMBH, whose frequency evolves. Consistent unimodal posteriors are obtained across all noise realisations and all five displayed parameters. The one-dimensional posteriors are biased with respect  to the injection values due to the unmodelled component of $\Omega_{\rm psr}^{(n)}$. Ten curves are plotted, corresponding to ten realisations of $\boldsymbol{Y}_{\rm evolving}$, but the curves overlap and are hard to distinguish by eye.}
	\label{fig:chi_chromatic}
\end{figure}

In this section we calculate the joint posterior probability distribution $p({\boldsymbol{\theta}'} | {\boldsymbol{Y}_{\rm evolving}})$. We consider two representative systems, analogously to Section \ref{sec:psr_v_earth_pe}. We again consider a ``low-SNR'' system with $h_0 = 5 \times 10^{-15}$ and a ``high-SNR'' system with $h_0 = 1 \times 10^{-12}$. All other static parameters are as specified in Table \ref{tab:parameters_and_priors}. \newline 

The results for $p({\boldsymbol{\theta}_{\rm gw}} | {\boldsymbol{Y}_{\rm evolving}})$ and $p({\boldsymbol{\theta}_{\rm gw}} | {\boldsymbol{Y}})$ are shown in Figure \ref{fig:monochromatic_assumption_high_snr}  for the high-SNR system and in Figure \ref{fig:monochromatic_assumption_low_snr} for the low-SNR system. The corner plots are arranged identically to Figures \ref{fig:corner_plot_compare_high} and \ref{fig:corner_plot_compare_low}, except that the different coloured curves correspond to posterior distributions inferred on different data. The blue curves are the results for $p({\boldsymbol{\theta}_{\rm gw}} | {\boldsymbol{Y}})$. The green curves are the results for $p({\boldsymbol{\theta}_{\rm gw}} | {\boldsymbol{Y}_{\rm evolving}})$. The axes cover a subset of the prior domain and are identical to the scales in Figures \ref{fig:corner_plot_compare_high} and \ref{fig:corner_plot_compare_low}, with the exception of the axis for $h_0$ in Figure \ref{fig:monochromatic_assumption_low_snr}, which covers a slightly broader range. \newline 

For the high-SNR results in Figure \ref{fig:monochromatic_assumption_high_snr}, the one-dimensional posteriors for $p({\boldsymbol{\theta}_{\rm gw}} | {\boldsymbol{Y}_{\rm evolving}})$ and $p({\boldsymbol{\theta}_{\rm gw}} | {\boldsymbol{Y}})$ are similar for five of the seven parameters in $\boldsymbol{\theta}_{\rm gw}$. The two exceptions are $\Phi_0$ and $h_0$. Regarding $\Phi_0$, the mode of the one-dimensional posterior
$p(\Phi_0 | {\boldsymbol{Y}_{\rm evolving}})$ is offset from the mode of $p(\Phi_0 | {\boldsymbol{Y}})$ by $\approx$ 0.4 radians. The shift can be understood by inspecting Equation \eqref{eq:g_func_trig_chi_chromatic}; the inference model based on Equation \eqref{eq:g_func_trig_chi} attempts to compensate for the 
unmodelled phase induced by $\Omega_{\rm psr}^{(n)}$ by adjusting the value of $\Phi_0$. The estimates of $\chi^{(n)}$ are similarly biased. In Figure \ref{fig:chi_chromatic} we plot the results for a representative subset for five out of 47 pulsar phases $ \chi^{(1)} \dots \chi^{(5)}$, for ten realisations of $\boldsymbol{Y}_{\rm evolving}$.
The figure is exactly analogous to Figure \ref{fig:corner_plot_3}. The one-dimensional histograms are evidently biased with respect to the injection value for the same reason as $\Phi_0$; the inference model must account for the unmodelled component of $\Omega_{\rm psr}^{(n)}$. Regarding $h_0$, the mode of the one dimensional posterior
$p(h_0 | {\boldsymbol{Y}_{\rm evolving}})$ is offset from the mode of $p(h_0 | {\boldsymbol{Y}})$ by $\approx$ $5 \times 10^{-13}$. Similarly to $\Phi_0$, this shift occurs due to the unmodelled phase component of $\Omega_{\rm psr}^{(n)}$, which manifests as a correction to the amplitude. The shift is comparable to the known dispersion (cosmic variance)in $h_0$ (cf. Appendix \ref{sec:app_high_SNR} and Figure \ref{fig:corner_high_snr_appendix}). \newline 

We define a relative error to quantify the accuracy of the one-dimensional posteriors with respect to the injected value, 
analogously to Section \ref{sec:psr_v_earth_pe} and Equation \eqref{eq:mean_rel_error}, viz.
\begin{eqnarray}
	\Delta_{I}(\theta) = \frac{ \left| \underset{\theta}{\text{argmax }} p_{\rm M}\left(\theta| I\right) - \theta_{\rm inj}\right|}{\theta_{\rm inj}} \, , \label{eq:mean_rel_error_2}
\end{eqnarray}
where the subscript $I \in \{ \boldsymbol{Y}, \boldsymbol{Y}_{\rm evolving} \}$ indicates whether the posterior is estimated using monochromatic data (i.e. synthetic data generated using Equation \eqref{eq:g_func_trig_chi}) or evolving data (i.e. synthetic data generated using Equation \eqref{eq:g_func_trig_chi_chromatic}), respectively. The error $\Delta_{I}(\theta)$ for each parameter in Figure \ref{fig:monochromatic_assumption_high_snr} is summarised in the upper half of Table \ref{tab:posterior_errors_chromatic}. \newline 

\begin{table}
	\centering
		\begin{tabular}{lcll}
			\toprule
			&$\theta$ & $\Delta_{\boldsymbol{Y}}(\theta)$ & $\Delta_{\boldsymbol{Y}_{\rm evolving}}(\theta)$  \\
			\hline
			\multirow{7}{2mm}{High SNR} & $\Omega$       &  $1.4 \times 10^{-5}$ & $9.0 \times 10^{-5}$\\
			& $\Phi_0$ & $1.8 \times 10^{-2}$ &$1.9 \times 10^{0}$ \\
			& $\psi$ &  $3.4 \times 10^{-4}$& $2.8 \times 10^{-3}$\\
			& $\iota$ &  $5.7 \times 10^{-2}$& $1.3 \times 10^{-1}$ \\
			& $\delta$ &  $5.9 \times 10^{-4}$& $1.5 \times 10^{-2}$ \\
			&$\alpha$ &  $2.2 \times 10^{-4}$ & $2.0 \times 10^{-3}$ \\
			&$h_0$ &  $4.0 \times 10^{-2}$ & $5.0 \times 10^{-1}$ \\
			\hline
			\multirow{7}{2mm}{Low SNR} & $\Omega$       & $7.8 \times 10^{-4}$ &$4.3 \times 10^{-3}$ \\
			& $\Phi_0$ &  $3.3 \times 10^{-1}$ & $1.0 \times 10^{0}$  \\
			& $\psi$ &  $7.7\times 10^{-2}$ & $1.4 \times 10^{-1}$ \\
			& $\iota$&  $3.0 \times 10^{-1}$& $1.1 \times 10^{-1}$ \\
			& $\delta$ &  $2.2 \times 10^{-1}$&  $1.6 \times 10^{-1}$ \\
			&$\alpha$ &  $2.0 \times 10^{-1}$& $3.1 \times 10^{-1}$\\
			&$h_0$ &  $3.4 \times 10^{-1}$ & $4.7 \times 10^{-1}$ \\
			\bottomrule
		\end{tabular}
		\caption{Same as Table \ref{tab:posterior_errors} but for relative error $\Delta_{I}(\theta)$, Equation \eqref{eq:mean_rel_error_2}, where the subscript $i \in \{ \boldsymbol{Y}, \boldsymbol{Y}_{\rm evolving} \}$ indicates whether the posterior is estimated using monochromatic data (i.e. synthetic data generated using Equation \eqref{eq:g_func_trig_chi}) or evolving data (i.e. synthetic data generated using Equation \eqref{eq:g_func_trig_chi_chromatic}), respectively. The top and bottom halves of the table record the high-SNR ($h_0 =1 \times 10^{-12}$) and low-SNR ($h_0 = 5 \times 10^{-15}$) tests respectively. The injected values are summarised in Table \ref{tab:parameters_and_priors}. For high SNR, we obtain $\Delta_{\boldsymbol{Y}}(\theta) < \Delta_{\boldsymbol{Y}_{\rm evolving}}(\theta)$ for $\theta \in \boldsymbol{\theta}_{\rm gw}$. For low-SNR, we obtain $\Delta_{\boldsymbol{Y}}(\theta) < \Delta_{\boldsymbol{Y}_{\rm evolving}}(\theta)$ for five out of seven parameters, although the discrepancy is modest.}
		\label{tab:posterior_errors_chromatic}
	\end{table}
	
	Table \ref{tab:posterior_errors_chromatic} confirms that we have $\Delta_{\boldsymbol{Y}}(\theta) < \Delta_{\boldsymbol{Y}_{\rm evolving}}(\theta)$ for $\theta \in \boldsymbol{\theta}_{\rm gw}$, i.e. the estimates are always more accurate when the inference is run on $\boldsymbol{Y}$. However, for five out of seven parameters the difference is modest; the parameters are recovered with high accuracy for $\boldsymbol{Y}$ and $\boldsymbol{Y}_{\rm evolving}$. For example, we find $| \Delta_{\boldsymbol{Y}}(\Omega) -  \Delta_{\boldsymbol{Y}_{\rm evolving}}(\Omega) | = 7.6 \times 10^{-5}$. For the remaining two parameters, $\Phi_0$ and $h_0$, the difference is greater; we find $\Delta_{\boldsymbol{Y}_{\rm evolving}}(\Phi_0)/  \Delta_{\boldsymbol{Y}}(\Phi_0)  \approx 10^2$
	and $\Delta_{\boldsymbol{Y}_{\rm evolving}}(h_0)/  \Delta_{\boldsymbol{Y}}(h_0)  \approx 10^1$. We present only a single noise realisation in this section, but the improvements from including the pulsar terms are found to be comparable across different realisations. \newline 
	
	For the low-SNR results in Figure \ref{fig:monochromatic_assumption_low_snr}, the one-dimensional posteriors for $p({\boldsymbol{\theta}_{\rm gw}} | {\boldsymbol{Y}_{\rm evolving}})$ and $p({\boldsymbol{\theta}_{\rm gw}} | {\boldsymbol{Y}})$ resemble each other for all seven parameters $\boldsymbol{\theta}_{\rm gw}$. Qualitatively, the green and blue contours and histograms mostly overlap, although the blue contours are centred slightly better on the injected values in some cases (e.g. $\Omega, \Phi_0$). Again, given the observed dispersion at low-SNR (e.g. Figure \ref{fig:corner_plot_1}), it is difficult to drawn strong conclusions about whether  $p({\boldsymbol{\theta}_{\rm gw}} | {\boldsymbol{Y}_{\rm evolving}})$ is more accurate than $p({\boldsymbol{\theta}_{\rm gw}} | {\boldsymbol{Y}})$. The relative error $\Delta_{I}(\theta)$ for the low-SNR results is reported in the lower half of Table \ref{tab:posterior_errors_chromatic}. Due to the reduced GW signal strength and the correspondingly broader posteriors,
	the high-SNR hierarchy $\Delta_{\boldsymbol{Y}}(\theta) < \Delta_{\boldsymbol{Y}_{\rm evolving}}(\theta)$ for $\theta \in \boldsymbol{\theta}_{\rm gw}$ no longer holds. In some cases (e.g. $\iota$, $\delta$), we have $\Delta_{\boldsymbol{Y}}(\theta) > \Delta_{\boldsymbol{Y}_{\rm evolving}}(\theta)$. However the difference is modest, and the posteriors inferred for $\boldsymbol{Y}$ and $\boldsymbol{Y}_{\rm evolving}$ are comparable.

	\subsection{Detection probability}\label{app:detection_chromatic}
	\begin{figure}
		\includegraphics[width=\columnwidth, height =\columnwidth ]{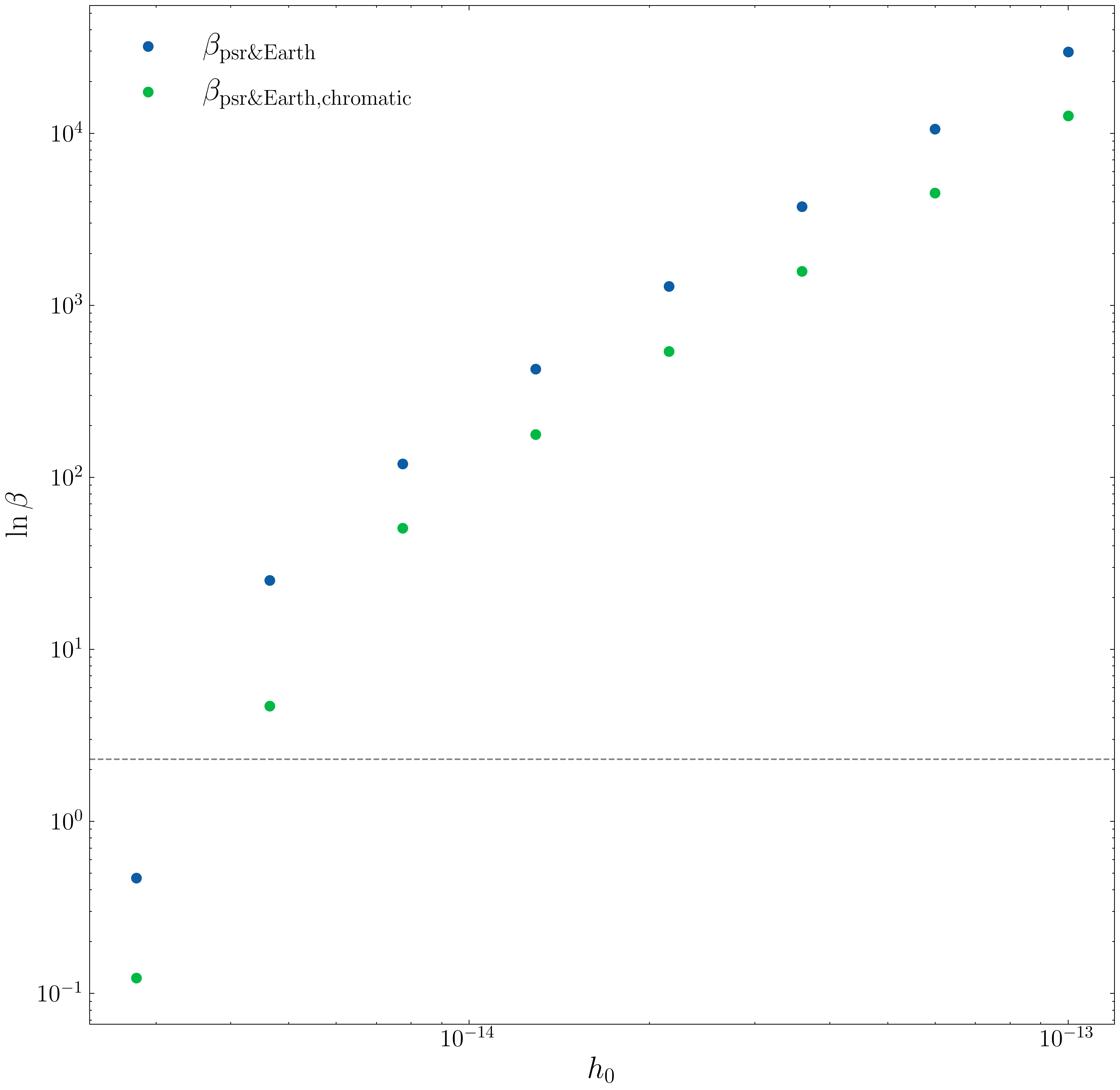}
		\caption{Bayes factors calculated using the inference model $\mathcal{M}_{\rm  \rm psr \& Earth}$ acting on $\boldsymbol{Y}$ ($\beta_{\rm psr{\&}Earth}$, blue points) and on $\boldsymbol{Y}_{\rm evolving}$ 
			($\beta_{\rm psr{\&}Earth, evolving}$, green points) as a function of the signal amplitude, $h_0$, for the representative example in Table \ref{tab:parameters_and_priors}, cf. Figure \ref{fig:bayes1}. The horizontal grey dashed line labels an arbitrary detection threshold, $\beta= 10$. The minimum detectable strain at $\beta= 10$ equals $2.9 \times 10^{-15}$ for $\boldsymbol{Y}$ and  $3.7 \times 10^{-15}$ for $\boldsymbol{Y}_{\rm evolving}$. The axes are plotted on logarithmic scales.}
		\label{fig:bayes_monochromatic_test}
	\end{figure}

	In this section we calculate the minimum detectable strain for the data $\boldsymbol{Y}_{\rm evolving}$ using the inference model $\mathcal{M}_{\rm psr \& Earth}$. We compare the minimum detectable strain to that calculated for the monochromatic data, $\boldsymbol{Y}$, considered in the main text. We follow Section \ref{sec:psr_v_earth_bayes} and frame the detection problem in terms of a Bayesian model selection procedure. We emphasise that we compare the minimum detectable strain for the same inference model acting on different data. In contrast, in Section \ref{sec:psr_v_earth_bayes} we compare the minimum detectable strain for different inference models acting on the same data. \newline

	The Bayes factors $\beta_{\rm psr \& Earth, evolving}$ and $\beta_{\rm psr \& Earth}$ are plotted as functions of $h_0$ in Figure \ref{fig:bayes_monochromatic_test}. We vary the source amplitude from $h_0 = 10^{-15}$ (undetectable) to $h_0 = 10^{-13}$ (easily detectable). As in Section  \ref{sec:psr_v_earth_bayes}  we present only a single noise realisation of the synthetic data pair $\boldsymbol{Y}$ and $\boldsymbol{Y}_{\rm evolving}$. The conclusions drawn below are consistent across different noise realisations. \newline 
	
	Figure \ref{fig:bayes_monochromatic_test} reveals $\beta_{\rm psr \& Earth} > \beta_{\rm psr \& Earth, evolving}$ for all $h_0$. We follow Section \ref{sec:psr_v_earth_bayes} and take $\beta = 10$ as an arbitrary detection threshold. For monochromatic data $\boldsymbol{Y}$, the GW source is detectable for $h_0 \gtrsim 2.9 \times 10^{-15}$. For data from evolving sources, $\boldsymbol{Y}_{\rm evolving}$,  the GW source is detectable for $h_0 \gtrsim 3.7 \times 10^{-15}$. That is, the minimal detectable strain deteriorates by 26\%. As in Section \ref{sec:psr_v_earth_bayes}, the minimum detectable strain is particular to the system in Table \ref{tab:parameters_and_priors} and the realisations of $\boldsymbol{Y}$ and $\boldsymbol{Y}_{\rm evolving}$. \newline

\section{Kalman filter} \label{sec:kalman}
The Kalman filter \citep{Kalman1} is an optimal solver for Gauss-Markov processes. Given a temporal sequence of noisy measurements, $\boldsymbol{Y}(t)$, the Kalman filter recovers a temporal sequence of stochastically evolving system state variables, $\boldsymbol{X}(t)$, which are hidden from the observer. It is a standard method in control theory \citep{zarchan2000fundamentals,rob_KF_book,10.5555/3103280} and finds common use in engineering applications \citep[e.g.][]{KFexample1,KFexample2,KFexample3,KFexample4} as well as successful application to neutron star astrophysics \citep[e.g.][]{Myers2021MNRAS.502.3113M,Meyers2021,Melatos2023}. In this appendix, we outline the Kalman filter used in this paper. The general recursion relations for the discrete-time Kalman filter are presented for an arbitrary linear dynamical system in Section \ref{sec_kalman_general}, along with the formula for the Bayesian likelihood. The application to the specific continuous-time state-space model in Section \ref{sec:2} is outlined in Section \ref{sec_kalman_specific}.

\subsection{Recursion equations and likelihood }\label{sec_kalman_general}
In this work we use the linear Kalman filter, which assumes linear relations between $d{\boldsymbol{X}}/dt$ and ${\boldsymbol{X}}(t)$ (dynamics) and between ${\boldsymbol{Y}}(t)$ and ${\boldsymbol{X}}(t)$ (measurement). The linear Kalman filter operates on temporally discrete, noisy measurements $\boldsymbol{Y}_k = \boldsymbol{Y}(t_k)$, which are related via a linear transformation to a set of unobservable discrete system states $\boldsymbol{X}_k= \boldsymbol{X}(t_k)$. Each discrete timestep is indexed by $ 1 \leq k  \leq K$. The measurements are related to the states via
\begin{equation}
	\boldsymbol{Y}_k = \boldsymbol{H}_k \boldsymbol{X}_k + \boldsymbol{v}_k \ ,\label{eq:kalman1}
\end{equation}
where $\boldsymbol{H}_k$ is the measurement matrix or observation model, $\boldsymbol{v}_k$ is a zero-mean Gaussian measurement noise, one has $\mathcal{N} \sim (0,\boldsymbol{R}_k)$ with covariance $\boldsymbol{R}_k$, and the subscript $k$ labels the time-step. The Kalman filter evolves the underlying states according to
\begin{equation}
	\boldsymbol{X}_k = \boldsymbol{F}_k \boldsymbol{X}_{k-1} + \boldsymbol{G}_k \boldsymbol{u}_k + \boldsymbol{w}_k \ , \label{eq:kalman2}
\end{equation}
where $\boldsymbol{F}_k$ is the system dynamics matrix, $\boldsymbol{G}_k$ is the control matrix. $\boldsymbol{u}_k$ is the control vector, and $\boldsymbol{w}_k$ is a zero-mean Gaussian process 
noise, with $\boldsymbol{w}_k \sim {\cal N}(0, \boldsymbol{Q}_k)$ and covariance $\boldsymbol{Q}_k$. \newline 

The Kalman filter is a recursive estimator with two distinct stages: a ``predict'' stage and an ``update'' stage. The predict stage predicts $\hat{\boldsymbol{X}}_{k|k-1}$, the estimate of the state at discrete step $k$, given the state estimate from step $k-1$. Specifically, the predict step proceeds as
\begin{align}
	\hat{\boldsymbol{X}}_{k|k-1} &=  \boldsymbol{F}_k \hat{\boldsymbol{X}}_{k-1|k-1} + \boldsymbol{G}_k \boldsymbol{u}_k \ , \\
	\hat{\boldsymbol{P}}_{k|k-1} &=  \boldsymbol{F}_k \hat{\boldsymbol{P}}_{k-1|k-1} \boldsymbol{F}_k^\intercal + \boldsymbol{Q}_k  \ ,
\end{align}
where $\hat{\boldsymbol{P}}_{k|k-1}$ is the covariance of the prediction. Note that the predict stage is independent of the measurements. The measurement $\boldsymbol{Y}_k$ updates the prediction during the update stage as follows:
\begin{align}
	\boldsymbol{\epsilon}_{k} &= \boldsymbol{Y}_k - \boldsymbol{H}_k \hat{\boldsymbol{X}}_{k|k-1} \ , \label{eq:residual} \\
	\boldsymbol{S}_k &= \boldsymbol{H}_k \hat{\boldsymbol{P}}_{k|k-1} \boldsymbol{H}_k^\intercal + \boldsymbol{R}_k \ , \label{eq:innv_covar}\\
	\boldsymbol{K}_k &= \hat{\boldsymbol{P}}_{k|k-1} \boldsymbol{H}_k^\intercal \boldsymbol{S}_k^{-1} \ ,\label{eq:kalman gain} \\
	\hat{\boldsymbol{X}}_{k|k} &=\hat{\boldsymbol{X}}_{k|k-1} +\boldsymbol{K}_k  \boldsymbol{\epsilon}_{k}  \ , \label{eq:kalmangainupdate} \\
	\hat{\boldsymbol{P}}_{k|k} &= \left( \boldsymbol{I} - \boldsymbol{K}_k \boldsymbol{H}_k \right) 	\hat{\boldsymbol{P}}_{k|k-1} \ .
\end{align}
Equation \eqref{eq:residual} defines a residual $\boldsymbol{\epsilon}_k = \boldsymbol{Y}_k  - \hat{\boldsymbol{Y}}_k$, which is sometimes termed the innovation. The uncertainty in $\boldsymbol{\epsilon}_k$ is quantified via the innovation covariance $\boldsymbol{S}_k = \langle \boldsymbol{\epsilon}_k \boldsymbol{\epsilon}_k^{T} \rangle$, viz. Equation \eqref{eq:innv_covar} (noting that the Einstein summation convention is suppressed temporarily in the latter definition). Equation \eqref{eq:kalman gain} defines the Kalman gain $\boldsymbol{K}_k$ For a full review of the Kalman filter, including its derivation, we refer the reader to \cite{Gelb:1974} and \cite{zarchan2000fundamentals}. \newline 

The Gaussian log-likelihood of obtaining ${\boldsymbol{Y}}_k$ given ${\boldsymbol{\hat X}}_k$ is calculated at each timestep from the Kalman filter output according to \citep{zarchan2000fundamentals}
\begin{eqnarray}
	\log \mathcal{L}_k =  -\frac{1}{2} \left (D_k \log 2 \pi + \log  \left | \boldsymbol{S}_k \right | + \boldsymbol{\epsilon}_k^{\intercal} \boldsymbol{S}_k^{-1}  \boldsymbol{\epsilon}_k \right ) \ ,
\end{eqnarray}
where $D_k$ is the dimension of $\boldsymbol{\epsilon}_k$ at timestep $k$. The total log-likelihood for the entire sequence is
\begin{eqnarray}
	\log \mathcal{L} =  \sum_{k=1}^{K} \log \mathcal{L}_k \ . \label{eq:likelihood}
\end{eqnarray}
Given ${\boldsymbol{Y}}_k$, $\mathcal{L}$ is a function of the estimates ${\boldsymbol{\hat \theta}}$ of the static parameters passed to the Kalman filter, i.e. $\mathcal{L}$ = $\mathcal{L}(\boldsymbol{Y} | \boldsymbol{\hat \theta})$. Similarly the estimates of the state and measurement variables, $\hat{\boldsymbol{X}}$ and $\hat{\boldsymbol{Y}}$, are functions of $\boldsymbol{\hat \theta}$. In Appendix \ref{sec:nested_sampling}, we explain how to combine the Kalman filter with a nested sampler to iteratively guide ${\boldsymbol{\hat \theta}}$ towards the true value of ${\boldsymbol{\theta}}$. \newline 

To apply the Kalman filter in practice means specifying the eight component matrices that make up the ``machinery'' of the filter: $\boldsymbol{X}_k$, $\boldsymbol{Y}_k$, $\boldsymbol{F}_k$, $\boldsymbol{G}_k$, $\boldsymbol{u}_k$, $\boldsymbol{H}_k$, $\boldsymbol{Q}_k$ and $\boldsymbol{R}_k$. In Appendix \ref{sec_kalman_specific} we define the machinery for the state-space model in Section \ref{sec:2}.

\subsection{State space representation of a PTA analysis}\label{sec_kalman_specific}
We apply the Kalman recursion relations in Section \ref{sec_kalman_general} to the state-space model of a PTA with $N$ pulsars described in Section \ref{sec:2} as follows. \newline

We identify $\boldsymbol{X}(t)$ with a vector of length $N$ composed of the intrinsic pulsar frequency states, i.e. 
\begin{equation}
	\boldsymbol{X}(t) = \left(f_{\rm p}^{(1)}(t), f_{\rm p}^{(2)}(t), ..., f_{\rm p}^{(N)}(t)\right) \ .
\end{equation}
Analogously,  we package the measured pulsar frequencies as
\begin{equation}
	\boldsymbol{Y}(t) = \left(f_{\rm m}^{(1)}(t), f_{\rm m}^{(2)}(t), ..., f_{\rm m}^{(N)}(t) \right) \ .
\end{equation}
The states evolve according to the continuous stochastic differential equation (c.f. Equation \eqref{eq:frequency_evolution})
\begin{equation}
	d \boldsymbol{X} = \boldsymbol{A} \boldsymbol{X} dt + \boldsymbol{C}(t) dt + \boldsymbol{\Sigma} d \boldsymbol{B}(t) \ , \label{eq:kalmn2}
\end{equation}
where $\boldsymbol{A}$ is a diagonal $N \times N$ matrix,
\begin{equation}
	\boldsymbol{A} = \text{diag} \left(-\gamma^{(1)}, -\gamma^{(2)}, ..., -\gamma^{(N)}\right) \ ,
\end{equation}
and $\boldsymbol{C}(t)$ is a time-dependent $N \times 1$ vector with $n$-th component
\begin{equation}
	C^{(n)}(t) =\gamma^{(n)} \left[ f^{(n)} _{\rm em} (t_1) + \dot{f}^{(n)} _{\rm em}(t_1) \, (t-t_1) \right] +  \dot{f}^{(n)} _{\rm em}(t_1)\ .
\end{equation}
The $N \times N$ matrix $\boldsymbol{\Sigma}$  governs the magnitude of the increments of Brownian motion (Wiener process) $d\boldsymbol{B}(t)$, with
\begin{equation}
	\boldsymbol{\Sigma} = \text{diag} \left(\sigma^{(1)}, \sigma^{(2)}, ..., \sigma^{(N)}\right) \ .
\end{equation}

In the idealized model in this paper, each pulsar's rotational state evolves phenomenologically according to a mean-reverting Ornstein-Uhlenbeck process, described by a Langevin equation, Equation \eqref{eq:kalmn2}, whose general solution is given by \citep{gardiner2009stochastic}
\begin{equation}
	\boldsymbol{X}(t) = e^{\boldsymbol{A} t} \boldsymbol{X}(0) + \int_0^t dt'  e^{\boldsymbol{A}(t-t')} \boldsymbol{C}(t') + \int_0^t d\boldsymbol{B}(t') e^{\boldsymbol{A}(t-t')} \boldsymbol{\Sigma}  \ . \label{eq:gardenier}
\end{equation} 
From Equation \eqref{eq:gardenier} we construct the discrete, recursive solution for $\boldsymbol{X}(t_k) = \boldsymbol{X}_k$ in the form of Equation \eqref{eq:kalman2}, with
\begin{align}
	\boldsymbol{F}_k &= e^{\boldsymbol{A} \Delta t } \  \\
	&= \text{diag}\left(e^{- \gamma^{(1)} \Delta t},e^{- \gamma^{(2)} \Delta t},...,e^{- \gamma^{(N)} \Delta t} \right) \ ,
\end{align}
\begin{align}
	\boldsymbol{G}_k \boldsymbol{u}_k &= \int_{t_k}^{t_{k+1}} dt' e^{\boldsymbol{A}\left( t_{k+1} - t' \right)}  \boldsymbol{C}(t')  \ , \\
	&= \left(G^{(1)}_k, G^{(2)}_k,...,G^{(N)}_k \right) ,
\end{align}
\begin{equation}
	\boldsymbol{w}_k = \int_{t_k}^{t_{k+1}} d \boldsymbol{B}(t') e^{\boldsymbol{A}\left( t_{k+1} - t' \right)} \boldsymbol{\Sigma}  \ ,  \label{eq:appendix_noise}
\end{equation}
\begin{align}
	G_k^{(n)} =&    f^{(n)}_{\rm em}(t_1) + \dot{f}^{(n)}_{\rm em}(t_1)  \left(\Delta t + t_k \right) \nonumber \\ 
	&- e^{-\gamma \Delta t} \left[  f^{(n)}_{\rm em}(t_1) +\dot{f}^{(n)}_{\rm em}(t_1)  t_k \right] \ ,
\end{align}
and $\Delta t = t_{k+1} - t_k$. From Equation \eqref{eq:appendix_noise} the process noise covariance matrix is
\begin{equation}
	\boldsymbol{Q}_k \boldsymbol{\delta}_{kj}= \langle \boldsymbol{\eta}_k \boldsymbol{\eta}_j^\intercal \rangle = \text{diag} \left(Q^{(1)}, Q^{(2)},...,Q^{(N)}\right) \ ,
\end{equation}
with 
\begin{equation}
Q^{(n)} = \frac{[\sigma^{n}]^2}{2 \gamma^{(n)}} \left[ 1 - e^{-2 \gamma^{(n)} \Delta t}\right] \ .
\end{equation}
The two remaining unspecified component matrices of the Kalman filter are the measurement matrix $\boldsymbol{H}_k$ and the measurement covariance matrix $\boldsymbol{R}_k$. These are defined straightforwardly from Equations \eqref{eq:measurement}--\eqref{eq:g_func_trig}. Specifically, 
$\boldsymbol{H}_k$ is a diagonal matrix where the $n$-th component of the diagonal is given by $g^{(n)}(t_k)$ from Equation \eqref{eq:g_func_trig}. The measurement covariance satisfies $\boldsymbol{R}_k = E \left[ \boldsymbol{v} \boldsymbol{v}^\intercal \right] = \sigma^2_{\rm m}$ for all $k$.

\section{Nested sampling}\label{sec:nested_sampling}
We can use the likelihood returned by the Kalman filter, Equation \eqref{eq:likelihood}, in conjunction with likelihood-based inference methods to estimate the posterior distribution of $\boldsymbol{\theta}$ by Bayes' Rule,
\begin{equation}
	p(\boldsymbol{\theta} | \boldsymbol{Y}) = \frac{\mathcal{L}(\boldsymbol{Y} | \boldsymbol{\theta}) \pi(\boldsymbol{\theta})}{\mathcal{Z}} \ ,
\end{equation}
where $\pi(\boldsymbol{\theta})$ is the prior distribution on $\boldsymbol{\theta}$ and $\mathcal{Z}$ is the marginalised likelihood, or evidence,
\begin{equation}
	\mathcal{Z} = \int d \boldsymbol{\theta} \mathcal{L}(\boldsymbol{Y} | \boldsymbol{\theta})  \pi(\boldsymbol{\theta})  \ . \label{eq:model_evidence2}
\end{equation}
We estimate the posterior distribution and the model evidence through nested sampling \citep{Skilling} in this paper. Nested sampling evaluates marginalised likelihood integrals, of the form given by Equation \eqref{eq:model_evidence2}. It also approximates the posterior by returning samples from $p(\boldsymbol{\theta} | \boldsymbol{Y})$. It does so by drawing a set of $n_{\rm live}$ live points from $\pi(\boldsymbol{\theta})$ and iteratively replacing the live point with the lowest likelihood with a new live point drawn from $\pi(\boldsymbol{\theta})$, where the new live point is required to have a higher likelihood than the discarded point. The primary advantage of nested sampling is its ability to compute $\mathcal{Z}$, on which model selection relies, as in Sections \ref{sec:pulsar_term} and \ref{sec:psr_v_earth_bayes}. Nested sampling is also computationally efficient and can handle multi-modal problems \citep{Ashton2022}. For these reasons, it has enjoyed widespread adoption in the physical sciences, particularly within the cosmological community \citep{Mukherjee2006,Feroz2008,Handley2015}, neutron star astrophysics \citep{Myers2021MNRAS.502.3113M,Meyers2021,Melatos2023}, particle physics \citep{proceedings2019033014} and materials science \citep{2009arXiv0906materials}. For reviews of nested sampling we refer the reader to \cite{Buchner2021} and \cite{Ashton2022}. Multiple nested sampling algorithms and computational libraries exist \citep[e.g.][]{Feroz2008,Feroz2009,Handley2015,dynesty2020,UltraNest2021}. In gravitational wave research it is common to use the \texttt{dynesty} sampler \citep{dynesty2020} via the \texttt{Bilby} \citep{bilby.507.2037A} front-end library. We follow this precedent and use \texttt{Bilby} for all nested sampling Bayesian inference in this work. \newline 

The primary tunable parameter in nested sampling is $n_{\rm live}$. More live points address larger parameter spaces and multi-modal problems, whilst the uncertainties in the evidence and the posterior scale as $\mathcal{O}\left(n_{\rm live}^{-1/2}\right)$. However the computational runtime scales as $\mathcal{O}(n_{\rm live})$. \cite{Ashton2022} offered a rule-of-thumb trade-off, where the minimum number of live points should be greater than the number of static parameters. Informal empirical tests conducted as part of this paper support the trade-off suggested by \cite{Ashton2022}; we find typically that the true ${\boldsymbol{\theta}}$ is contained within the 90\% credible interval of the one-dimensional marginalised posteriors of ${\boldsymbol{\hat{\theta}}}$ for $n_{\rm live} > 7 + 5N$ with $N \leq 50$. Unless stated otherwise we take $n_{\rm live} = 2000$ conservatively for all results presented in this work. \newline

\section{Workflow summary}\label{sec:workflow}
For the reader's convenience we now summarise the workflow for a representative PTA analysis using the Kalman filter and nested sampler for parameter estimation and model selection:
\begin{enumerate}[leftmargin=2em]
	\item Specify a PTA composed of $N$ pulsars 
	\item Obtain $N$ data inputs $f_{\rm m}^{(n)}(t)$, collectively labelled $\boldsymbol{Y}$
	\item Specify a state-space model $\mathcal{M}$, with static parameters $\boldsymbol{\theta}$
	\item Specify prior distribution $\pi(\boldsymbol{\theta})$
	\item Sample $n_{\rm live}$ points from $\pi(\boldsymbol{\theta})$ 
	\item For each live point:
	\begin{enumerate}[leftmargin=2em]
		\item Pass the sample $\boldsymbol{\theta}_{\rm sample}$ to the Kalman filter
		\item Iterate over the input data using the Kalman filter and obtain a single $\log \mathcal{L}$ value through Equation \eqref{eq:likelihood}
	\end{enumerate}
	\item Remove the live point with the lowest likelihood value, $\log \mathcal{L}_{\rm lowest}$
	\item Sample a new live point from $\pi(\boldsymbol{\theta})$, subject to the requirement that the new likelihood obeys $\mathcal{L}_{\rm new}$ > $\mathcal{L}_{\rm lowest}$, where $\log \mathcal{L}_{\rm new}$ is calculated via steps (vi)(a)--(vi)(b).
	\item Update $p\left(\boldsymbol{\theta}|\boldsymbol{Y}\right)$ and $\mathcal{Z}$ with nested sampler
	\item Repeat steps (vii)--(ix) until convergence criteria are satisfied.
\end{enumerate}
In order to compute the odds ratio $\beta$ the above workflow is repeated for a different $\mathcal{M}$. The resulting $\mathcal{Z}$ values can then be divided. We remind the reader that the above workflow differs from a realistic PTA analysis in one important respect, namely that the data are input as frequency time series $f_{\rm m}^{(n)}(t)$ instead of pulse TOAs. The generalization to TOAs is subtle and will be tackled in a forthcoming paper.

\section{Challenges of the pulsar term}\label{sec:psr_term_challenges}

\begin{figure*}
	\includegraphics[width=\textwidth, height =0.5\textwidth ]{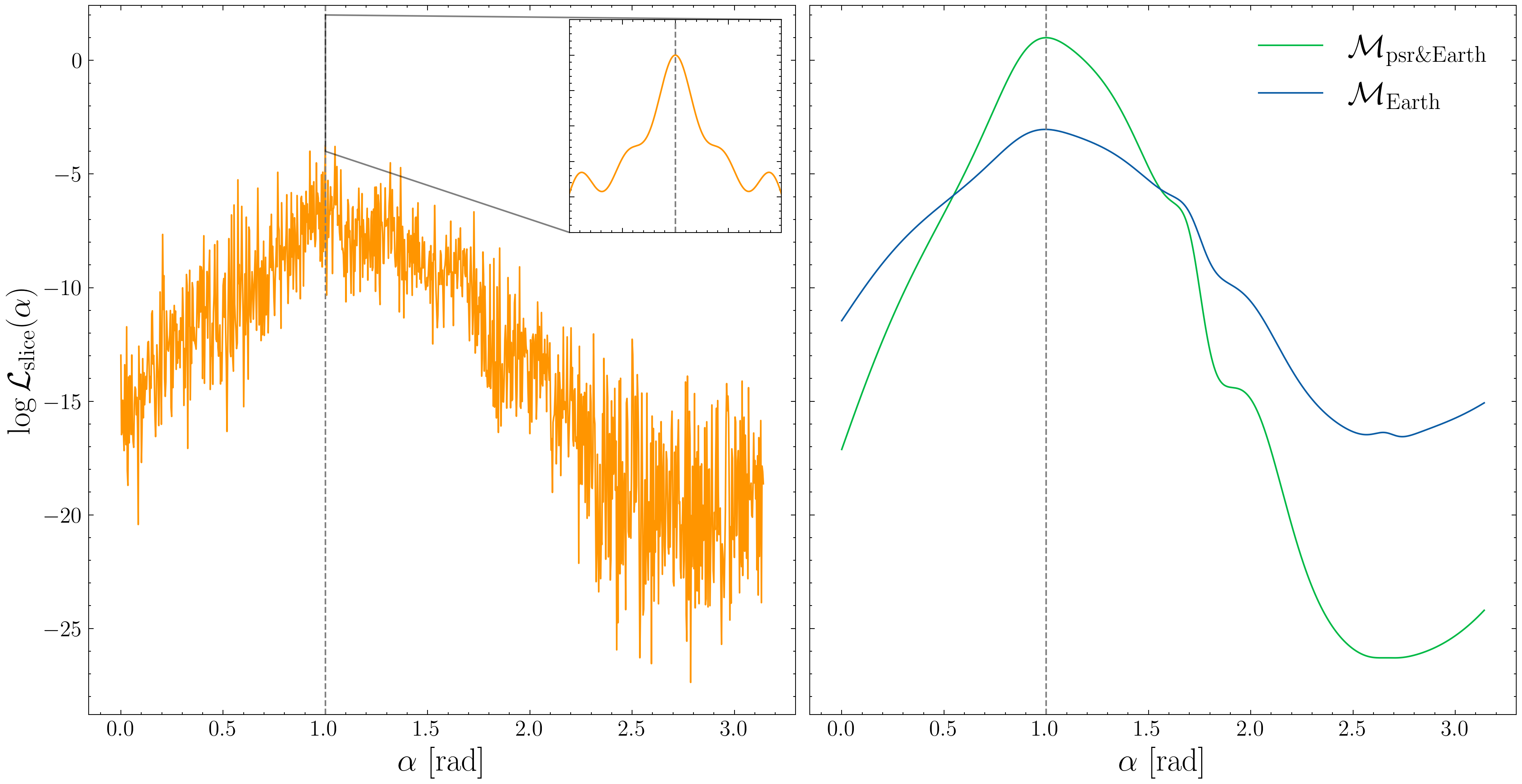}
	\caption{Logarithm of the cross-section along the $\alpha$-axis of the likelihood function $\mathcal{L}_{\rm slice}(\alpha)$ returned by the Kalman filter acting on synthetic data $\boldsymbol{Y}$. All static elements of $\boldsymbol{\theta}$ other than $\alpha$ are held fixed at their true injected values for the purpose of the tests in Appendix \ref{sec:psr_term_challenges}. In the left panel the Kalman filter uses Equation \eqref{eq:g_func_trig} to calculate $\mathcal{L}_{\rm slice}(\alpha)$, i.e. inclusive of the pulsar terms. In the right-hand panel the Kalman filter uses Equation \eqref{eq:g_func_trig_earth}, i.e. just the Earth terms. The vertical dashed lines indicate the true injected value $\alpha = 1.0$ rad used to generate $\boldsymbol{Y}$. The inset in the left panel covers the region $1.0 - 10^{-4} \le \alpha \le 1.0 +10^{-4}$ rad. In the left panel $\mathcal{L}_{\rm slice}(\alpha)$ is jagged, although a unique peak exists in principle at $\alpha=1.0 \, {\rm rad}$, and $\mathcal{L}_{\rm slice}(\alpha)$ is smooth within the neighbourhood plotted in the inset. The jagged form of $\mathcal{L}_{\rm slice}(\alpha)$ hampers the convergence of the nested sampler. In the right panel $\mathcal{L}_{\rm slice}(\alpha)$ is smooth for $0\leq\alpha\leq \pi$. The green (blue) curve is the likelihood returned by $\mathcal{M}_{\rm psr\&Earth}$ ($\mathcal{M}_{\rm Earth}$).}
	\label{fig:likelihood_comparison}
\end{figure*}
In this appendix we elucidate the complications introduced by the pulsar terms into the state-space analysis and justify the reparametrisation in terms of $\chi^{(n)}$ introduced in Section \ref{sec:pulsar_term2}. \newline 

As a first step, we demonstrate empirically that sticking with the original parameterization of the pulsar terms in Equation \eqref{eq:g_func_trig}, i.e.\ writing them in terms of $d^{(n)}$ instead of $\chi^{(n)}$, produces a highly oscillatory likelihood function, which the nested sampler fails to navigate successfully. Physically, this happens because the Doppler factor depends on the GW wavelength, and there are $\gtrsim 10^3$ GW wavelengths across the Galactic volume spanned by the PTA. To enact the demonstration, we generate noisy synthetic data $\boldsymbol{Y}$ as outlined in Section \ref{sec:rep_example1}. The SMBHB source parameters $\boldsymbol{\theta}_{\rm gw}$ are chosen arbitrarily, with $\alpha = 1.0 $ rad. We run the Kalman filter using the measurement equation inclusive of the pulsar terms, Equation \eqref{eq:g_func_trig}, for $ 0 \le \alpha \le \pi$. All other elements of $\boldsymbol{\theta}_{\rm gw}$ are fixed at their true injected values. At this point we do not use the $\chi^{(n)}$ parametrisation. Therefore changing $\alpha$ also changes $\boldsymbol{n}$ everywhere it appears in Equation \eqref{eq:g_func_trig}. For each value of $\alpha$, the Kalman filter returns a likelihood $\mathcal{L}(\boldsymbol{Y} | \alpha) = \mathcal{L}_{\rm slice}(\alpha)$, i.e.\ a cross-section of $\mathcal {L}(\boldsymbol{\theta}_{\rm gw})$ along the $\alpha$-axis for the purpose of testing. \newline

The function $\mathcal{L}_{\rm slice}(\alpha)$ is displayed in the left panel of Figure \ref{fig:likelihood_comparison} (orange curve). The dashed grey line indicates the injected value $\alpha= 1.0$ rad. The inset shows a zoomed-in section of $\mathcal{L}(\alpha)$ within the region $ 1.0 - 10^{-4} \le \alpha \le 1.0 +10^{-4}$ rad.  There are two important features in the left panel. First, on scales of the order the width of the prior $\pi(\alpha)$ (i.e radians), the curve $\mathcal{L}_{\rm slice}(\alpha)$ is jagged. Similar jagged curves are obtained for $\mathcal{L}_{\rm slice}(\Omega)$ and $\mathcal{L}_{\rm slice}(\delta)$ (not plotted for brevity). For the other static parameters, the curves are smooth and globally concave. Second, despite being jagged, a true likelihood maximum coincident with the injected value does exist and is locally concave in a neighbourhood spanning $\sim 10^{-4} \, {\rm rad}$. Hence it is theoretically possible to use likelihood-based methods for Bayesian estimation of $\alpha$ without reparameterizing in terms of $\chi^{(n)}$. In practice, however, the computation is intractable and does not converge for reasonable choices of $n_{\rm live}$. The challenge for sampling algorithms is exacerbated when we extend the above test from $\mathcal{L}_{\rm slice}(\alpha)$ to the $\alpha$-$\delta$-$\Omega$ subspace or the full $7 + 5N$ domain in Section \ref{sec:ss_params}. \newline 

Why do $\alpha$, $\delta$ and $\Omega$ exhibit jagged likelihood curves whereas other static parameters do not? The culprit is the Doppler factor flagged above: $\alpha$, $\delta$, and $\Omega$ appear in the phase term $\Omega \left[1 + \boldsymbol{n}\cdot \boldsymbol{q}^{(n)} \right]  d^{(n)}$, with $\boldsymbol{n} = \boldsymbol{n}(\delta, \alpha)$. For $d^{(n)} \sim 1$ kpc, a 10 nHz GW accumulates $ \gtrsim 10^3$ cycles. The inference problem becomes multiply degenerate, once the cosine of the phase is calculated (modulo $2\pi$). \newline

In order to smooth the likelihood function, one can try two things. The first is to drop the pulsar term completely, i.e. using $\mathcal{M}_{\rm Earth}$. This is the approach taken in K24 and some (not all) other standard PTA analyses \citep[e.g.][]{Sesana2010,Babak2012,Petiteau2013,Zhu2015,Taylors2016,Goldstein2018,Charisi2023arXiv230403786C}. The second is to reparameterize the pulsar terms such that the phase term $\Omega \left[1 + \boldsymbol{n}\cdot \boldsymbol{q}^{(n)} \right]  d^{(n)}$ is no longer a function of $\alpha$ (or $\Omega$ or $\delta$), but an independent parameter $\chi^{(n)}$ to be estimated. This is the approach presented in this paper, i.e. using $\mathcal{M}_{\rm psr \& Earth}$, as in Section \ref{sec:pulsar_term2}. The right panel of Figure \ref{fig:likelihood_comparison} displays $\mathcal{L}_{\rm slice}(\alpha)$ calculated using $\mathcal{M}_{\rm Earth}$ (blue curve) and $\mathcal{M}_{\rm psr \& Earth}$ (green curve). In both instances there is no ambiguity about the number of cycles the wave has gone through over a distance $d^{(n)}$, and the likelihood function is smooth.

\section{Priors}\label{sec:set_priors}
To deploy the nested sampling outlined in Appendix \ref{sec:nested_sampling} and the workflow in Appendix \ref{sec:workflow} it is necessary to specify a Bayesian prior $\pi(\boldsymbol{\theta})$ on the static parameters. In this section we outline how the priors are chosen. The priors on each static parameter are summarised in Table \ref{tab:parameters_and_priors}.  \newline 

For $\pi(\boldsymbol{\theta}_{\rm gw})$ we assume no \textit{a priori} information about the parameters. We therefore choose standard non-informative priors \citep[e.g.][]{Bhagwat2021}. For $\pi(\boldsymbol{\theta}_{\rm psr})$ \textit{a priori} information from electromagnetic observations does exist. We adopt constrained uniform priors on $f_{\rm em}^{(n)}(t_1)$ and $\dot{f}_{\rm em}^{(n)}(t_1)$, which extend $\pm 10^3 \eta_f^{(n)}$ and $\pm 10^3 \eta_{\dot{f}}^{(n)}$ respectively about the central, injected values, where $\eta_f^{(n)}$ and $\eta_{\dot{f}}^{(n)}$ denote the errors quoted in the ATNF Pulsar Database. By using wider-than-necessary priors we expose the analysis scheme to a more stringent test. We set an uninformative broad prior $\pi[\sigma^{(n)} / (1 \, {\rm s^{-3/2}})] \sim$ LogUniform$ \left [10^{-2} \sigma_{\rm SC}^{(n)}, 10^2 \sigma_{\rm SC}^{(n)} \right ]$
where $\sigma_{\rm SC}^{(n)}$ is the noise amplitude for pulsar $n$ inferred from
Equation \eqref{eq:sigmap_f}. We set a non-informative prior on the new static parameter $\chi^{(n)} \sim \text{Uniform}\left(0,2\pi\right)$. We do not set a prior on $\gamma^{(n)}$, because one typically has $\gamma^{(n)} T_{\rm obs} \sim 10^{-5}$ astrophysically, and $\gamma^{(n)}$ is effectively ``unobservable'' for $T_{\rm obs} \sim 10 \, {\rm years}$. For validation purposes it is sufficient to carry $\gamma^{(n)}$ through the analysis at its injected value. This reduces the total dimension of the parameter space to $7 + 4N$. \newline

\section{High-SNR example} \label{sec:app_high_SNR}
In this appendix we repeat the analysis of Section \ref{sec:pe_and_ms} in the high-SNR regime. We do so as a sanity check to confirm that the state-space analysis scheme with the pulsar terms included returns reasonable results in the ``easy'' case with a strong GW signal. The high-SNR limit allows us to check for estimation biases unobscured by measurement noise. \newline

We apply the parameter estimation framework as in Section \ref{sec:pe_and_ms}. All injected static parameters $\boldsymbol{\theta}'$, and the associated priors are specified in Table \ref{tab:parameters_and_priors}, with the exception of $h_0$ which now takes the value $h_0 = 1 \times 10^{-12}$. The procedure is undertaken for ten realisations of the noise processes $\xi^{(n)}(t)$ and $\varepsilon^{(n)}(t)$. \newline

\begin{figure*}
	\includegraphics[width=\textwidth, height = \textwidth ]{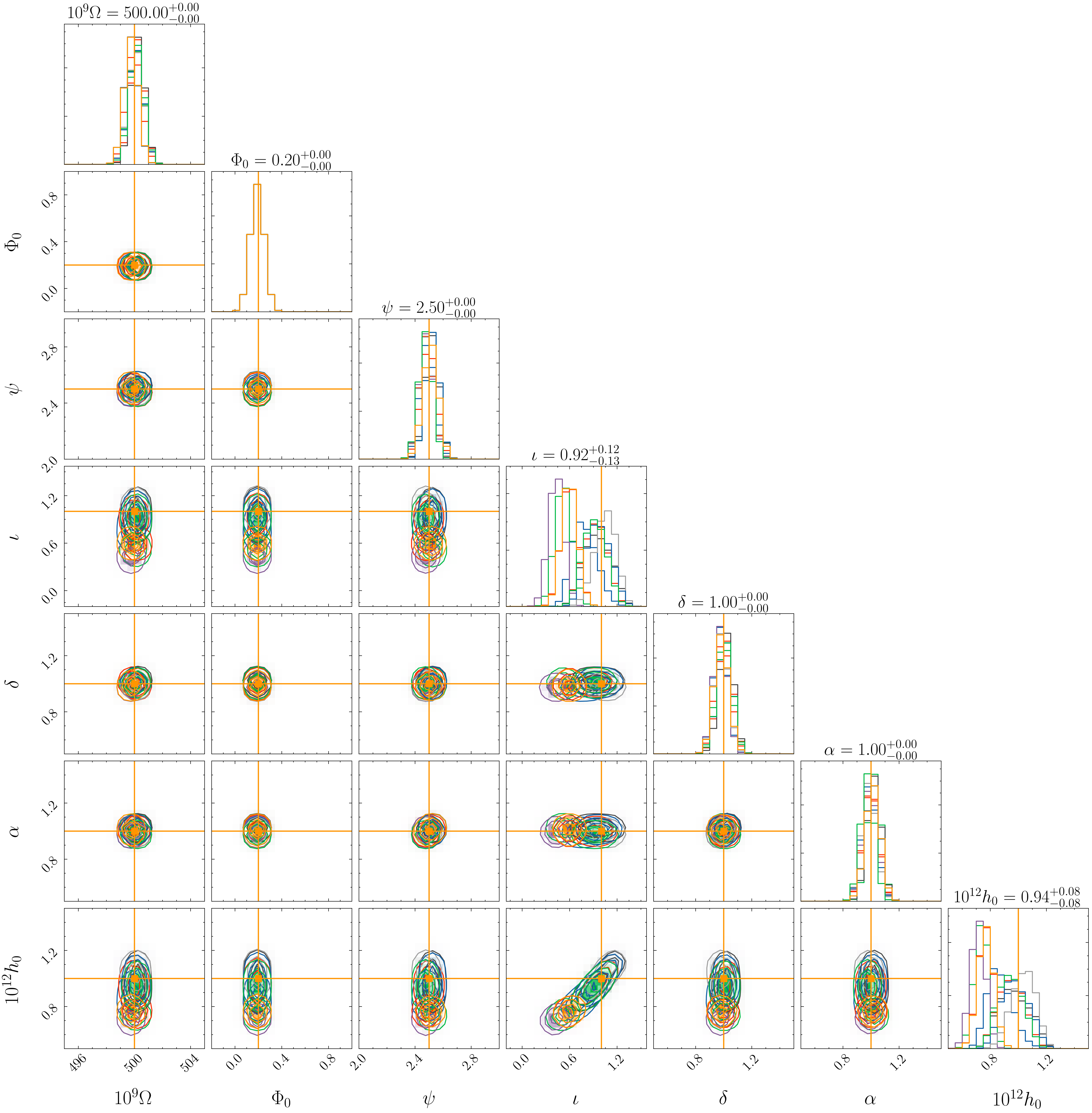} 	
	\caption{Same as Figure \ref{fig:corner_plot_1}, but for a loud SMBHB with $h_0 = 1 \times 10^{-12}$. The Kalman filter and nested sampler estimate all seven parameters in $\boldsymbol{\theta}_{\rm gw}$ accurately. There is less dispersion in the  one-dimensional posterior medians among noise realisations compared to the low-SNR case (c.f. Figure \ref{fig:corner_plot_1}). An appreciable dispersion remains for $\iota$ and $h_0$, which are correlated and weakly identifiable.}
	\label{fig:corner_high_snr_appendix}
\end{figure*}

Figure \ref{fig:corner_high_snr_appendix} displays results for the seven parameters in $\boldsymbol{\theta}_{\rm gw}$ for ten arbitrary noise realisations in the form of a traditional corner plot. The figure is exactly analogous to Figure \ref{fig:corner_plot_1}. All histograms and contours are consistent with a unimodal joint posterior, which peaks near the known, injected values. There is no evidence of railing against the prior bounds. Correlations between $\iota$ and $h_0$ are evident over multiple noise realisations due to the weak identifiability \footnote{By weak identifiability in this context we mean that the $\iota$-$h_0$ likelihood contours plateau to a flat-topped ridge with a small, non-zero gradient close to the maximum likelihood solution.} of these two parameters \citep{e5be7c83a0d24500826f6e1b414d1733}. For five of the seven parameters, there is effectively zero dispersion in the one-dimensional posterior medians between noise realisations. For the remaining two parameters, $\iota$ and $h_0$, appreciable dispersion is observed, again due to weak-identifiability. \newline

\begin{figure}
	\includegraphics[width=\columnwidth, height = \columnwidth ]{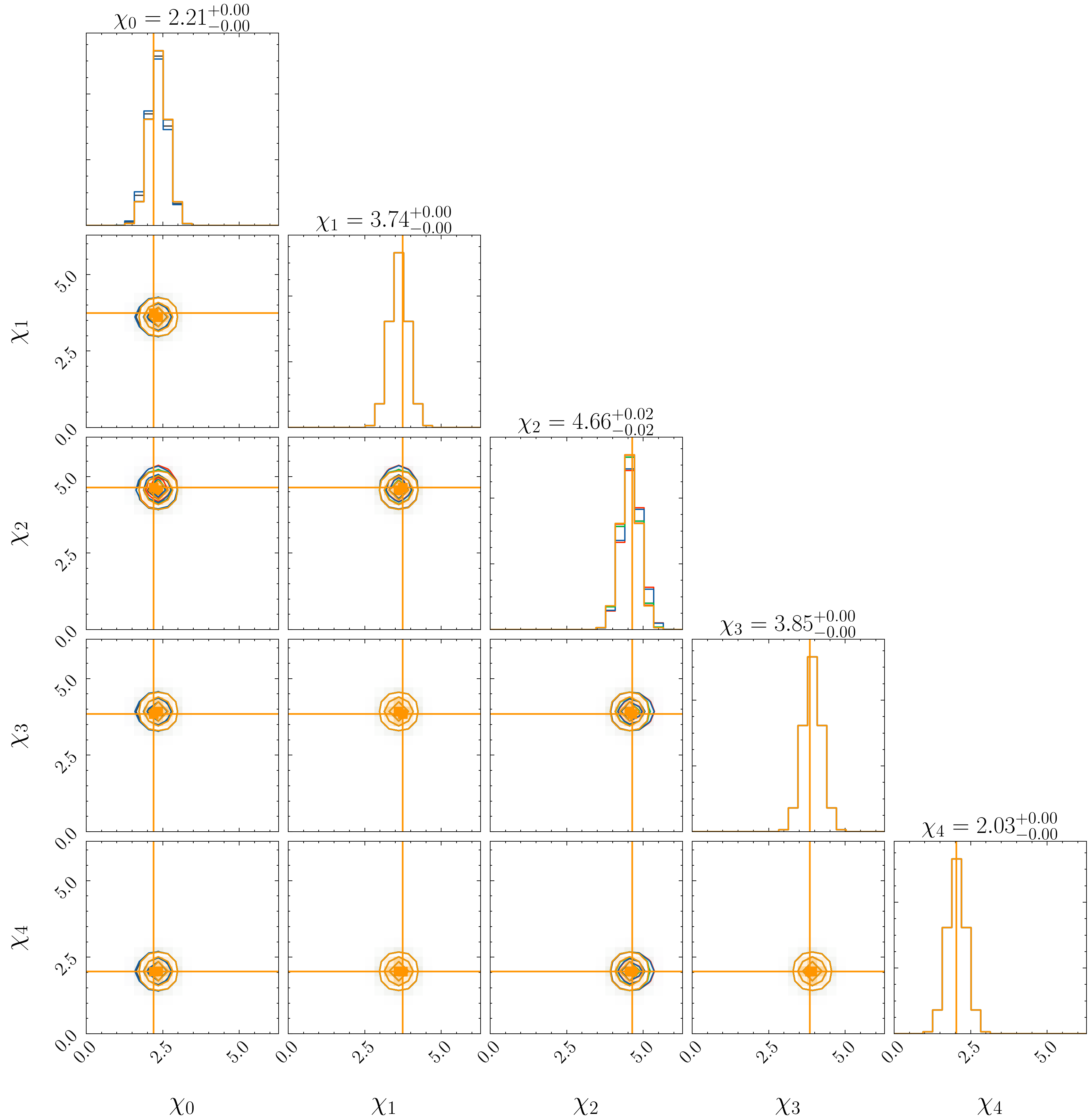} 	
	\caption{Same as Figure \ref{fig:corner_plot_3}, but for a loud SMBHB with $1 \times h_0 = 10^{-12}$.  Consistent unimodal posteriors are obtained across all noise realisations and all five displayed parameters. It is now possible to infer $\chi^{(2)}$ due to the loundness of the GW signal, unlike in the low-SNR case in Figure \ref{fig:corner_plot_3}. The colour scheme is the same as in Figure \ref{fig:corner_plot_3}, but all the curves overlap and so are obscured by the orange curves.}
	\label{fig:corner_high_snr_appendix_chi}
\end{figure}

Figure \ref{fig:corner_high_snr_appendix_chi} displays results for a representative subset of five out of 47 pulsar phases $\chi^{(n)}$, $ \{\chi^{(1)} \dots \chi^{(5)} \}$ for the ten noise realisations. The figure is exactly analogous to Figure \ref{fig:corner_plot_3}. All histograms and contours are consistent with a unimodal joint posterior, which peaks near the known, injected values, with no evidence for correlations between parameter pairs. Unlike in the low-SNR regime, it is now possible to infer $\chi^{(2)}$ (the phase correction for PSR J0340+4130) consistently (third row, third column of Figure \ref{fig:corner_high_snr_appendix_chi}). Even though PSR J0340+4130 nearly coincides on the sky with the synthetic GW source, the GW signal is loud enough for $\chi^{(2)}$ to be estimated.

\bsp	
\label{lastpage}
\end{document}